\documentclass[final,12pt]{elsarticle}

\usepackage{amssymb}
\usepackage{graphicx}
\usepackage{algorithm}
\usepackage{algpseudocode}
\usepackage{url}

\newcommand{\uhelp}{\mbox{uHelp}}

\begin{document}

\begin{frontmatter}



\title{uHelp: intelligent volunteer search for mutual help communities}


\author{Nardine Osman\corref{ca}}\ead{nardine@iiia.csic.es}
\author{Bruno Rosell}\ead{rosell@iiia.csic.es}
\author{Carles Sierra}\ead{sierra@iiia.csic.es}
\author{Marco Schorlemmer}\ead{marco@iiia.csic.es}
\author{Jordi Sabater-Mir}\ead{jsabater@iiia.csic.es}
\author{Lissette Lemus}\ead{lissette@iiia.csic.es}
\address{Artificial Intelligence Research Institute (IIIA-CSIC), Bellaterra, Spain}

%\cortext[ca]{corresponding author.}

\begin{abstract}
%When people need help with day-to-day tasks they turn to family, friends or neighbours to help them out. Finding someone to help out can be a stressful waste of time, especially when the task is sensitive and urgent, like finding someone to pick up one's child from school. Despite an increasingly networked world, technology falls short in supporting such daily irritations. \uhelp\ provides a platform for building a community of helpful people and supports them in finding help for their day-to-day tasks. \uhelp\ relies on a trio of techniques --- semantic similarity, a trust model, and a flooding algorithm --- to help efficiently find the most trusted volunteers in one's own social network for a given task request. This article presents the \uhelp\ application, describing the underlying AI technologies that allow \uhelp\ find most suitable volunteers efficiently, and illustrates the implementation details. \uhelp's initial prototype has been tested with a community of single parents in Barcelona, and the app is available online at both Apple Store and Google Play.
When people need help with their day-to-day activities, they turn to family, friends or neighbours.  
But despite an increasingly networked world, technology falls short in finding suitable volunteers.  
In this paper, we propose \uhelp, a platform for building a community of helpful people and supporting community members find the appropriate help within their social network.  Lately, applications that focus on finding volunteers have started to appear, such as Helpin or Facebook's Community Help. 
However, what distinguishes \uhelp\ from existing applications is its trust-based intelligent search for volunteers. 
Although trust is crucial to these innovative social applications, none of them have seriously achieved yet a trust-building solution such as that of \uhelp.  
\uhelp's intelligent search for volunteers is based on a number of AI technologies: (1) a novel trust-based flooding algorithm that navigates one's social network looking for appropriate trustworthy volunteers; (2) a novel trust model that maintains the trustworthiness of peers by learning from their similar past experiences; and (3) a semantic similarity model that assesses the similarity of experiences. %And it is this intelligent search that sets \uhelp\ apart from existing innovative social applications.
This article presents the \uhelp\ application, describes the underlying AI technologies that allow \uhelp\ find trustworthy volunteers efficiently, and illustrates the implementation details. \uhelp's initial prototype has been tested with a community of single parents in Barcelona, and the app is available online at both Apple Store and Google Play.

\end{abstract}

\begin{keyword}
social networks, flooding algorithm, trust model, semantic similarity, mobile app, communities of mutual help



\end{keyword}

\end{frontmatter}


\section{Introduction}\label{sec:intro}
The web has evolved into a social space enabling individuals to be pulled opportunistically into peer communities to achieve both personal and group goals.  This has been possible due to the widespread adoption of software applications that support and facilitate basic group interaction and collaboration such as file sharing, blogging, social networking, crowd funding, and crowd sourcing.   

But, despite the success of certain well-designed social software platforms such as Facebook, Twitter, or Whatsapp, there is still a need for a social platform that addresses users' call for help with day-to-day activities. In many cases, finding help with one's day-to-day activities can be crucial. For example, finding someone to pick up the children from school when one is running late at work, finding someone to pick up prescriptions and medicines for an elderly, or finding someone to substitute for a single parent at work when she needs to stay home with her sick child.

According to Beech et al.~\cite{Beech2004}, being late to pick up the children is a major stress factor for working parents.  From one of the interviews they conducted, they quote a mother: ``the worst time is the afternoons, and trying to finish off work to leave on time to collect my son from the nursery.''  A follow-up study~\cite{Sellen2004} indicates that one of the most severe problems that working parents encounter is coping with unexpected scheduling issues.  Support for parents is even more critical in nontraditional families, such as divorced or never married single-parent families. Marks and McLanahan~\cite{10.2307/352817} states that ``it is friends, not other kin, who figure most prominently in the social support relationships of nontraditional families", highlighting the need to focus on one's network of friends when in need.  
% NAR XXX some statistics on the elderly 

In this article, we present \uhelp\ \cite{uhelpAT,uhelpAAMAS}, a software application that provides a % NAR XXX fully distributed 
platform for building and maintaining a local community of people helping one another with their day-to-day tasks, thus supporting the balancing of societal needs, which contributes to community well-being. The \uhelp\ application essentially allows community members to search for appropriate volunteers within their community that would help them with their day-to-day activities. % NAR XXX add reference to the Springer uhelp chapter? 
This situates \uhelp\ as a typical application of the collaborative economy model. This is an economy model that has been described as ``the peer-to-peer-based activity of obtaining, giving, or sharing the access to goods and services, coordinated through community-based online services''~\cite{ASI:ASI23552}, and it has been growing in such a way that Forbes has estimated that ``revenue flowing through the sharing economy directly into people's wallets will surpass \$3.5 billion, with growth exceeding 25\%''~\cite{forbes2013}.

There have been a few platforms developed lately that focus on finding volunteers, such as Helpin,\footnote{\url{http://helpinapp.net}} an app for people to publicise what they need or what they can offer, and Facebook's Community Help,\footnote{\url{https://newsroom.fb.com/news/2016/11/facebooks-social-good-forum/}} a tool developed by Facebook that allows users to help each other after a disaster.  

However, the main novelty of \uhelp\ which distinguishes it from existing social network applications (whether it was generic applications, such as Twitter, Facebook, or Whatsapp, or more specific applications focused on volunteering, such as Helpin and Facebook's Community Help) is its intelligent search for volunteers. Existing social network platforms allow one to broadcast their request either publicly or privately for their group of friends. They may even allow one to create a temporary group of manually handpicked friends for discussing a certain topic (as in the case with Whatsapp). 

\uhelp\ has been designed for the more sensitive and urgent tasks, where one might not be interested in broadcasting a request, nor will it have the time to handpick its trusted friends for a given sensitive request, such as finding a volunteer to picking up one's child from school in half an hour.\footnote{Although \uhelp\ was designed with the more sensitive and urgent tasks it mind, it adequately handles other general cases as well.} Instead, the user will simply state its request, specify the rules of who can be trusted for this specific request, and press the help button. In fact, as Figures~\ref{fig:helpgeneral}--\ref{fig:helpprovide} illustrate, all of this is achieved in a few seconds with a very small number of clicks.  Specifying the requirements on who can be trusted for a specific request depends on two parameters: (1) the level of trustworthiness, and (2) the friendship level.  As an example, for picking up one's child from school, a mother may require only friends with utmost trustworthiness to be asked. For a similar request, another parent may accept friends of friends. However, to get the medication delivered, an elderly person may loosen the trust and friendship requirements further.  As such when making a request, the user is asked to specify the trust and friendship levels. 

Unlike any other social network platform, \uhelp\ crawls one's own social network for suitable volunteers. It does not only ask trusted friends, but when possible, may go looking for trusted peers further down the friendship line. It essentially propagates the request along one's own social network through a flooding algorithm that forwards a request from one node in one's social network to another based on the satisfaction of the required levels of trust and friendship.

As for measuring how much does one person trust another, we note that we do not require one to manually specify his trust on each of his friends (although this is possible), but we permit users to rate volunteers and then we use a trust model to update the trustworthiness of people based on learning from similar past experiences. In other words, the trust model calculates one's trust on another by predicting the quality of the other's future performance based on his past performance on the same or similar tasks. The trust model then relies on a semantic similarity module that would assess the similarity of past experiences. 

In summary, the main contribution of this article is \uhelp's intelligent search for volunteers, which is based on a combination of AI technologies: (1) a novel trust-based flooding algorithm that navigates one's social network looking for appropriate trustworthy volunteers; (2) a novel trust model that maintains the trustworthiness of peers by learning from similar past experiences; and (3) a semantic similarity model that assesses the similarity of experiences. And it is this intelligent search that sets \uhelp\ apart from existing innovative social applications. The importance of trust for creating social stability has been advocated over and over again. ``Trust is what makes contracts, plans and everyday transactions possible; it facilitates the democratic process, from voting to law creation, and is necessary for social stability. It is essential for our lives. It is trust, more than money, that makes the world go round."\cite{nyt2013} Although trust is crucial to collaborative economy, none of the existing applications (to date) achieve a trust-building solution such as that of \uhelp.  \uhelp\ essentially aims to build social capital through amplifying the user's immediate circle of trust, and it does that using a unique trio of AI techniques.

The remainder of this article is divided as follows. Section~\ref{sec:technologies} presents the main technologies behind the intelligent search for volunteers; Section~\ref{sec:implementation} illustrates our implementation of the \uhelp\ app; Section~\ref{sec:usecase} discusses our initial test of \uhelp\ where we tested it with a community of single mothers; and Section~\ref{sec:conclusion} closes with some concluding remarks.

\section{Intelligent search for volunteers}\label{sec:technologies}
As illustrated above, \uhelp's novelty and its main contribution to the social network platforms lies is its intelligent search for volunteers, which relies on a trio of models and mechanisms. Namely, a flooding algorithm that propagates one's own social network looking for trusted volunteers, a trust model that updates one's trust over another by learning from similar past experiences, and a semantic similarity model that supports the trust model by assessing which experiences are considered similar and to what degree. These three technologies are presented next: the flooding algorithm in Section~\ref{sec:flooding}, the trust model in Section~\ref{sec:trust}, and the semantic similarity model in Section~\ref{sec:semantics}

\subsection{Flooding algorithm}\label{sec:flooding}
The flooding algorithm is the core computational process in the \uhelp\ platform. It ensures requests for help are properly disseminated through the community. As stated earlier, we build upon a social network representation of the community: this is represented by some graph in which the members of the community are nodes and the edges represent friendship relations between two people. When someone wants to request help for a specific task, the flooding algorithm sends this request to that person's \emph{trusted} neighbours in the graph (i.e. its friends) and from there it continues to flood through the network to other trusted nodes. This is similar to a number of other algorithms designed for rapidly disseminating a message through a graph, most prominently the Gnutella algorithm for P2P file sharing. The main difference between existing approaches and the flooding algorithm discussed here is that the decision to stop forwarding the request is made primarily based on trust, rather than on other things, like the time passed since the initial request was made. In fact, in Figures~\ref{fig:helpgeneral}--\ref{fig:helpprovide}, one can see how the user controls the flooding algorithm by deciding the required trust level for a given task, along with the friendship level, which represents the number of hops in a graph (friends stand for 1 hop, friends of friends stand for 2 hops, etc.).

The reason for relying on trust is that we not only want to find someone willing to volunteer for a task, but the person must also be trustworthy when performing this task. The way we calculate trust between two people is described in Section~\ref{sec:trust}. To calculate trust along a path, we assume that trust satisfies the triangular norm inequality, that is for all $\alpha, \beta, \gamma$ nodes in a network $\mathit{Trust}(\alpha, \gamma) \le \mathit{T}(\mathit{Trust}(\alpha, \beta),\mathit{Trust}(\beta,\gamma))$ for some  T-norm function $T$. Thus, trust is monotonically decreasing along any path. We use the minimum for our experiments: $\mathit{Trust}(\alpha, \gamma) = \min\{\mathit{Trust}(\alpha, \beta), \mathit{Trust}(\beta,\gamma)\}$. Though other functions may be investigated, like multiplication ($\cdot$). The flooding algorithm stops propagating the request when the cumulative trustworthiness of a node falls below a certain threshold $\tau$. Algorithm \ref{flooding_algo_code} provides the pseudocode for the flooding algorithm.

\begin{algorithm}[!p]
\caption{Flooding algorithm}
\label{flooding_algo_code}
\begin{algorithmic}[1]\small
\Require $me : Node$ \Comment{The current node's identifier, of type $Node$.}
\Require $n \to m$ \Comment{The request to execute method $m$ at node $n$. Methods are defined as functions and are non-blocking.}
\Require $\mathit{friends} : 2^{Node}$ \Comment{The current node's set of neighbouring nodes.}
\Require $\mathit{Trust} : Node \times Task \to [0,1] \cup \bot$ \Comment{The current node's trust on some node for a given task. $\bot$ for unknown people.}
\Require $\mathit{T}: [0,1] \times [0,1] \to [0,1]$ \Comment{A T-norm function, e.g. $min$, $\cdot$.} \Require $\mathit{OldPathTrust} : Task \to [0,1]$ \Comment{Previous trust received from a path for a given tasks. Initially it is -1.}
\Require $ReceivedRequests : 2^{Task}$ \Comment{The current node's set of received requests.}
\Require $\sigma : [0,1]$ \Comment{Minimum increase in trust required to re-flood the network.}
\Require $Now() : Time$ \Comment{The current date/time.}
\Require $length : Path \to \mathbb{N}$ \Comment{A function that returns the length of a path.}
\Require $\oplus : Path \times Node \to Path$ \Comment{A function that appends a node to a path.}
\Require $Msg\_Help$, $Msg\_NotNeeded$, $Msg\_Cancelled$ \Comment{Functions that trigger popup messages, as illustrated in Section~\ref{sec:fsm}.}

\Function{Propagate}{$task$, $messagetype$, $\tau$, $pathtrust$, $path$, $deadline$, $hops$}
	\If{$me \not\in path$ and  $Now() < deadline$}\label{line:condition}
    	\If{$messagetype == \mathit{HELP}$}
			\If{$task \not\in ReceivedRequests$}\label{line:db}
				\State $\mathit{OldPathTrust}(task) := pathtrust$;
				\State $me$ $\to$ {\sc Flood}($task$, $messagetype$, $\tau$, $pathtrust$, $path\oplus me$, $deadline$, $hops$,  $\mathit{false}$);
			\ElsIf{$pathtrust - \mathit{OldPathTrust}(task) > \sigma$} \label{line:reflood} %or $newstatus \neq StatusRequestee(task)$}
				\State $\mathit{OldPathTrust}(task) := pathtrust$;
				\State $me$ $\to$ {\sc Flood}($task$, $messagetype$, $\tau$, $pathtrust$, $path\oplus me$, $deadline$, $hops$, $true$);
			\EndIf
    	\Else 
        	\State $me$ $\to$ {\sc Flood}($task$, $messagetype$, $\tau$, $pathtrust$, $path\oplus me$, $deadline$, $hops$, $\mathit{false}$);
        \EndIf
	\EndIf
%    	\State $StatusRequestee(task) := newstatus$;
\EndFunction

\Function{Flood}{$task$, $messagetype$, $\tau$, $pathtrust$, $path$, $deadline$, $hops$, $asked?$}
	\If{$\neg \ asked?$}
\algstore{myalg}
\end{algorithmic}
\end{algorithm}

\begin{algorithm}                     
\begin{algorithmic} [1]                   
\algrestore{myalg}
        \If{$messagetype == \mathit{HELP}$} 
			\State $me \to$ Msg\_Help($task$,$deadline$);
        \ElsIf{$messagetype == \mathit{NOTNEEDED}$}
        	\State $me \to$ Msg\_NotNeeded($task$);  
        \ElsIf{$messagetype == \mathit{CANCELLED}$}
        	\State $me \to$ Msg\_Cancelled($task$);
		\EndIf                 
	\EndIf
    %        \State $me$ $\to$ $msg(task,messagetype) $
	\ForAll{$n\in \mathit{friends}$}
		\State $\mathit{NewPathTrust} := \mathit{T}(\mathit{Trust}(me,n, task),pathtrust)$; \label{line:trust}
		\If{$\mathit{NewPathTrust} \geq \tau$ and $length(path) -1 < hops$}
			\State $n \rightarrow$ {\sc Propagate}($task$, $messagetype$, $\tau$, $\mathit{NewPathTrust}$, $path$, $deadline$,$hops$);
		\EndIf
	\EndFor
\EndFunction
\end{algorithmic}
\end{algorithm}

Every node in the network runs Algorithm~\ref{flooding_algo_code}. It works in a straightforward manner. Every time a user needs help, she fills in the help form and presses the help button (Figures~\ref{fig:helpgeneral}--\ref{fig:helpprovide}). Pressing the help button automatically calls the {\sc Flood} function, passing some information obtained from the form as arguments to the {\sc Flood} function. These are: 
\begin{itemize}
\item the $task$ details, such as the child to be picked up, the pickup location and time, etc.
\item the minimum level of trust $\tau$ required from potential volunteers (specified as the ``Trust'' field in Figures~\ref{fig:helpgeneral}--\ref{fig:helpprovide}); 
\item the maximum number of $hops$ permitted when flooding (specified as the ``Friendship'' field in Figures~\ref{fig:helpgeneral}--\ref{fig:helpprovide}); and 
\item the $deadline$ for responding to a request, which is calculated based on the task's deadline (specified for instance as the ``Deadline'' field in Figure~\ref{fig:helpprovide}). We note that the deadline for responding to a request should precede the task's deadline by a predefined fixed amount of time. For instance, if the deadline for the task is tonight at 6pm, then the deadline to accept may be today at 3pm, which is 3 hours before the task's deadline.
\end{itemize}

The other parameters of the {\sc Flood} function are automatically set accordingly. 
\begin{itemize} 
\item We specify the message type as $messagetype=\mathit{HELP}$. This is needed because the flooding algorithm is not only used to ask for help. For instance, when the user cancels a request before selecting a volunteer, then all those that have been asked for help should be informed that the request is now cancelled. In this case, we run the same flooding algorithm, but with $messagetype=\mathit{CANCELLED}$. Also, when a user selects a volunteer, then all those that have been asked for help (except the selected volunteer) should be informed that their help is no longer needed. In this case, we again run the same flooding algorithm, but with $messagetype=\mathit{NOTNEEDED}$. These cases are revisited in Section~\ref{sec:fsm}.
\item We set the initial trust accordingly: $pathtrust=1$. Recall that every time a message propagates to its neighbouring nodes, trust is calculated using the T-norm function ($T$). 
% NAR XXX multiplied by the trust on that new node. 
We start with the value 1, the neutral element of $T$. % NAR XXX That is, if the requester $r$ propagates the request to its neighbour $n$, then $r$'s trust on $n$ is taken as is initially multiplied by 1.  
\item We set the path of the people that this request has been delivered to accordingly: $path=\langle me\rangle$. In other words, the requester adds itself to the list, ensuring that the flooding algorithm does not later on ask the requester for help for his own request (see Line~\ref{line:condition} of Algorithm~\ref{flooding_algo_code}).
\item We set $asked?=yes$, which is a parameter that decides whether the user should be informed of the request or not (through the pop up message asking if the user is willing to volunteer). This parameter is needed because the node may receive and propagate the same request more than once. Yet, the human user should not be asked more than once whether it is willing to volunteer or not. We note here that we do allow a node to receive and propagate a message more than once, which we refer to as re-flooding, if received from different paths and certain other conditions are met. Re-flooding is discussed and motivated shortly.
\end{itemize}

In what follows we describe how the flooding algorithm works. When the {\sc Flood} function is triggered at a node, it first checks whether the human user should be informed of the corresponding request. If the user needs to be informed (i.e., if $\neg asked?$), then the appropriate pop-up message is triggered (which also adds the request to the requests list, as illustrated in Section~\ref{sec:requesteeFSM}). The request is then propagated to the node's neighbours, only when the requirements are fulfilled. The requirements are considered fulfilled when the trust on a neighbour is higher than the threshold $\tau$ and the number of hops is smaller than the acceptable limit $hops$. Recall that when adding a new node to the path, trust is updated by the T-norm function by aggregating %multiplying 
the trust on the node propagating the request by the trust the propagating node has on the propagated node (Line~\ref{line:trust} of Algorithm~\ref{flooding_algo_code}). 

A request is propagated by having the neighbouring nodes, that fulfil the requirements mentioned above, execute the {\sc Propogate} function. With the exception of excluding the $asked?$ parameter, the function has the same parameters as the {\sc Flood} function, though the $pathtrust$ is updated with the new trust value.  When a node executes the {\sc Propagate} function, it first verifies that there is no loop with the flooding (by checking that the node is not already included in the path) and that we are not beyond the deadline (Line~\ref{line:condition} of Algorithm~\ref{flooding_algo_code}). %Then, for help messages, the node first checks whether it has been asked for help. 
Then, the node checks whether it has been asked for help (Line~\ref{line:db} in Algorithm~\ref{flooding_algo_code}). 
If it has not, then the {\sc Flood} function is called, with the $asked?$ parameter set to $\mathit{false}$ and the node appending itself to the $path$. If it has been asked earlier for help for this specific task, then it is allowed to re-flood the network if the task has been requested previously but the path trust on the node is now sufficiently higher than before (by a difference larger than $\sigma$) (Line\ref{line:reflood} of Algorithm~\ref{flooding_algo_code}). In this case, the {\sc Flood} function is called, with the $asked?$ parameter set to $true$ and the node appending itself to the $path$. This last point is very important as the depth by which a request percolates the network depends on the cumulated level of trust. There may be several paths connecting any two nodes, and thus the trust between them should be considered as the maximum of the cumulated trust over the paths. If we reach a node with a certain level of trust we will flood the network with that value. However, if later on a higher value is found we need to re-flood the network as this high value may make nodes previously not reached by the non-increasing effect of the T-norm be reachable now.\footnote{We note that when the T-norm function is the minimum ($\min$), then re-flooding is not needed, and in fact does not occur, which is the case of our implementation. This is because with the minimum, the trust of a new path can only go down, but never up. As such, re-flooding never happens. However, for other implementations of the T-norm function, such as the product ($\cdot$), re-flooding may occur and is needed.}

Concerning efficiency, we note that the flooding algorithm is not affected by loops in the graph as flooding is stopped when the node itself is found in the request path (due to the condition on Line~\ref{line:condition} of Algorithm \ref{flooding_algo_code}).  Setting parameter $\sigma$, which helps decide when to re-flood (Line~\ref{line:reflood} of Algorithm~\ref{flooding_algo_code}), is also very important for the efficiency of the algorithm. If it is set very low then the number of messages can become exponential in the number of nodes for particular topologies. In the worst case, with $\sigma = 0$, for a graph $G$, the number of messages is $\sum_{p \in loop\_\mathit{free}\_paths(G)} length(p)$,\footnote{$loop\_\mathit{free}\_paths(G)$ returns the set of paths in a graph $G$ that do not contain any loops, and $length(p)$ returns the length of a path $p$.} which can be exponential in the number of nodes.  An adequate value has to be set experimentally.  
In comparison with other algorithms, we note that our algorithm limits the number of hops in addition to considering the trust level. This is achieved by comparing the length of the $path$ with the $hops$ parameter representing the maximum number of hops in the {\sc Flood} method.  This limitation in the number of hops makes our algorithm somehow similar to the Gnutella flooding algorithm, although in our case we also take trust into account. 

% NAR XXX TO MOVE
% XXX
% Finally, we note that because \uhelp\ is designed to run on mobile devices, it is important to take breaks in connectivity into account. In the implemented version of the algorithm, we make use of time limits, after which it is assumed that any node that has not answered was unreachable. This allows the algorithm to deal with users whose cellphones are switched off, or are unreachable for other reasons. Additionally, at a lower level, the messaging protocol could detect such failures and it can then resend the message when the receiving node is back online.
% NAR XXX time limits not implemented

\subsection{Semantic similarity}\label{sec:semantics}
The flooding algorithm is designed to flood to trustworthy nodes. We say the trustworthiness of a node is a measure that is learnt from the node's past performance in similar past experiences. For instance, if one has been rated highly in the past when donating electronic gadgets then she might be considered trustworthy enough to lend her speakers for a neighbourhood party. But before we present our proposed trust model, this section presents the semantic similarity model that assesses the similarities of experiences. 

The tasks requested by \uhelp\ users consist of the \emph{activity} the task is about (such as care for a relative, which could include picking up, babysitting, cooking, etc.). Activities have other relevant \emph{object} information, such as the \emph{relative} in the caring for a relative activity (Figure~\ref{fig:helpgeneral}), or the \emph{item} in the providing/donating activity (Figure~\ref{fig:helpprovide}). In this section, we will use the caring for a relative as an example of an activity and the relative as an example of an activity's object. And we will particularly focus on children, and hence we may have baby, toddler, preschooler, etc. All of this is just to keep the sample ontologies simple and clear. These examples also illustrate the need to handle two separate hierarchies.

\begin{figure}[!ht]
\centering
\includegraphics[width=\textwidth]{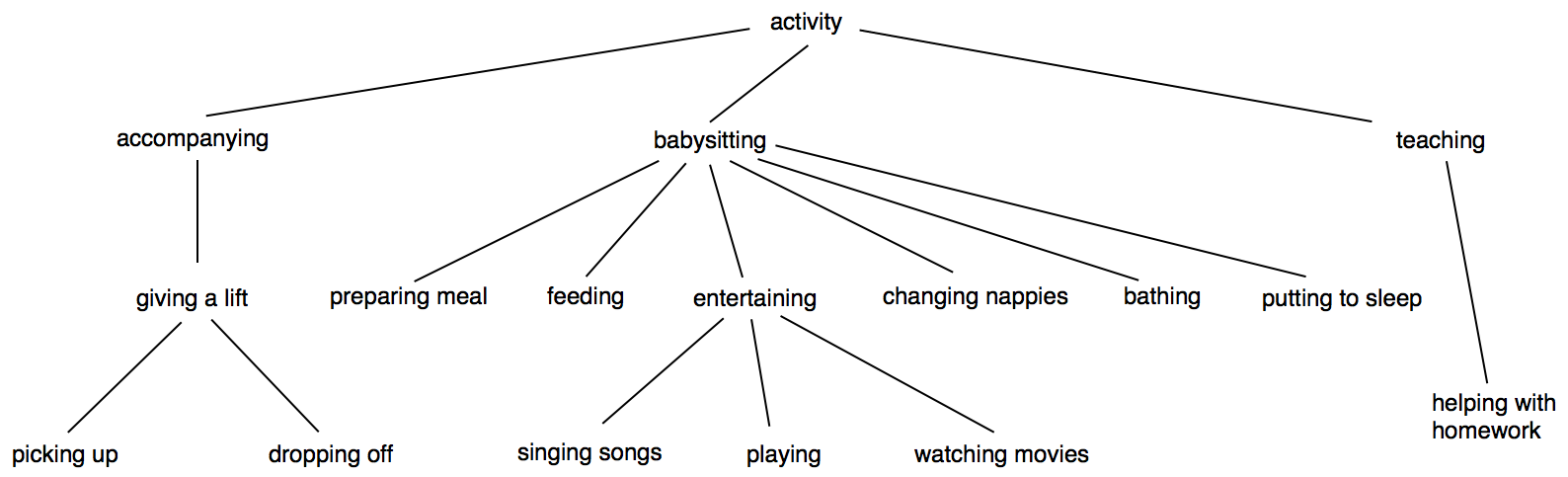}
\caption{Meronomy of an activity; in this case, focusing on providing care}
\label{fig:activity-meronomy}
\end{figure}

On one hand, the different kinds of activities are usually organised in a meronomy,\footnote{A meronomy is a type of hierarchy that deals with part-of relationships, in contrast to a taxonomy whose structure is based on is-a relationships.} specifying which sub-activities may be part of a given activity (e.g., changing nappies is a sub-activity of babysitting). Here we take inspiration from the design decisions suggested by the Process Specification Language (PSL)~\cite{Gruninger:2003:PSL:958671.958677}. An example of a meronomy of activities and sub-activities for the \uhelp\ caring for a relative task is given in Figure~\ref{fig:activity-meronomy}.

\begin{figure}[!ht]
\centering
\includegraphics[width=0.8\textwidth]{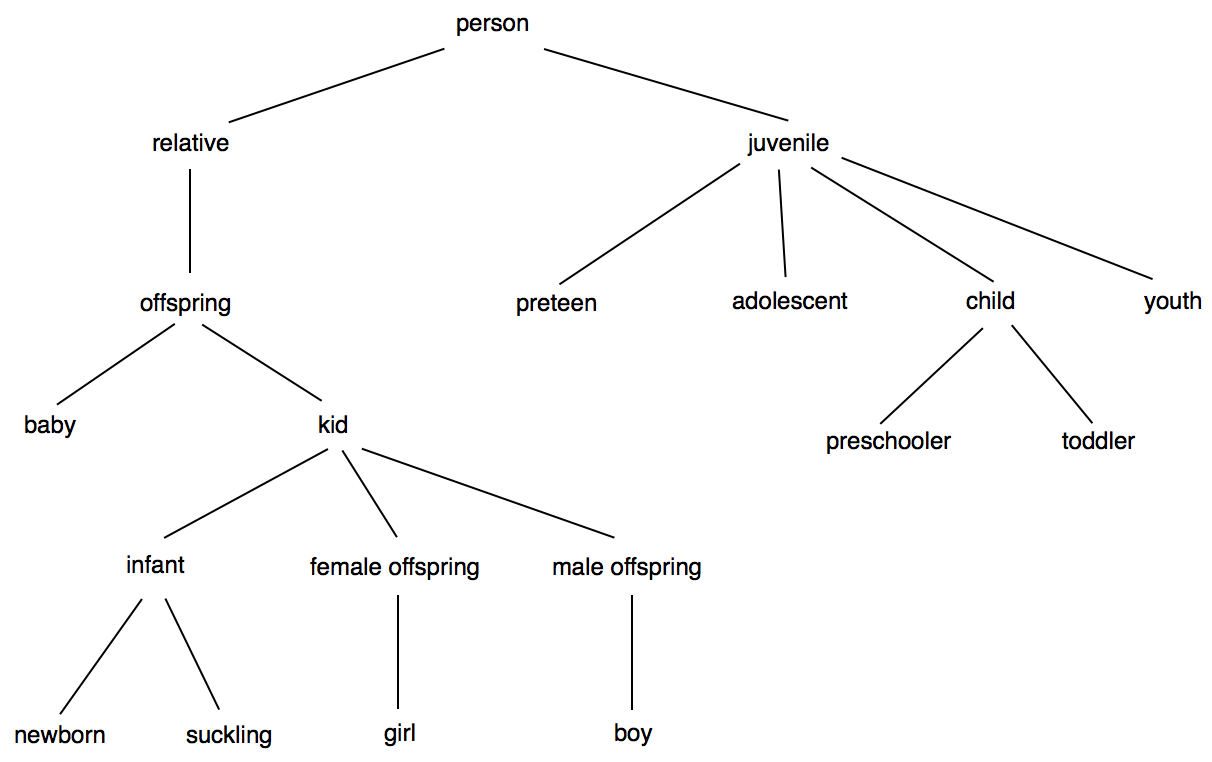}
\caption{Taxonomy of objects; in this case, focusing on children}
\label{fig:child-taxonomy}
\end{figure}

On the other hand, the different kinds of objects, children in this case, may sometimes be more naturally organised in a taxonomy. For instance, specifying subclasses of children: e.g., a toddler \emph{is a} child.
Figure~\ref{fig:child-taxonomy} shows an example of a taxonomy of children for our example that reflects a hypernym-hyponym relation, which is similar to the one defined in WordNet~\cite{fellbaum1998wordnet}.

The task becomes a paring of an activity and an object. For our specific example, this would be a pairing of a ``care for'' activity with a ``child'' object, for instance $\langle$\emph{giving a lift, preschooler}$\rangle$ or $\langle$\emph{feeding, toddler}$\rangle$. We note that if the activity specified is not a leaf of the meronomy, the requester expects the volunteer to be capable of performing all the subactivities involved. For instance, if the activity is \emph{giving a lift}, the volunteer is expected to perform both \emph{picking up} and \emph{dropping off}. However, in a taxonomy this is treated differently: if the child is not a leaf of the taxonomy it is simply not as detailed a specification: a schoolchild is either a preadolescent or an adolescent, but clearly not both.


This gives the requester the freedom to specify a task at a more, or less, detailed level.\footnote{Our current \uhelp\ application features six pre-defined tasks and one general task. In this implementation, the user is free to specify the level of detail of the task for general tasks only.} Naturally, the level of detail of the task can affect how the flooding algorithm propagates the request. %, since a potential volunteer's trustworthiness is calculated with respect to the specified task. 
When considering propagating a request to a potential volunteer, we assess the potential volunteer's trusworthiness with respect to the request (or task) in question. The \emph{similarity} between the requested task and the potential volunteer's previously performed tasks is necessary for estimating his performance by learning from his similar past experiences.\footnote{Note that trustworthiness is  specified as an estimation of future performance. That is, if the probability of one performing well in a future task is high, then one will be considered trustworthy for this task.}

%Consequently, task similarity has to reflect this influence on the quality of task performance. 
Given that a task is composed of an activity and its object, activities and their meronomical structure as well as objects and their taxonomical structure become crucial for trust evaluations. 
We deal with the similarity of activities directly by the trust model, by using OpinioNet \cite{Osman2010}, which is a algorithm that propagates ratings in a meronomy. In other words, if we know one's opinion about another's babysitting performance, then we can predict their opinion about the other's feeding or putting a child to sleep performances. OpinioNet and how it is used is discussed in more detail in Section~\ref{sec:trust}. 

However, we also argue that assessing one's performance with respect to an activity may depend a lot on the type of object as well. For instance, preparing a meal for a baby is very different from preparing a meal for an adolescent. We therefore need to take the similarity between objects, children in this case, into account as well.

The similarity over the object taxonomy is considered from a more semantic perspective and we use the measure proposed by Li et al.~\cite{Li2003}, which combines well-known edge-based and node-based techniques, and correlates well with similarity assessments as done by humans. In particular, it takes three aspects of taxonomies into account:
\begin{itemize}
\item the distance $l$ between two tasks $t_1$ and $t_2$ in the taxonomy: the closer, the more similar the tasks are;
\item the depth $h$ in the taxonomy of the most specific subsumer $sub(t_1,t_2)$ of two tasks $t_1$ and $t_2$: the deeper in the taxonomy the subsumer is, the more similar the tasks are;
\item the local semantic density $d$ of instances of these tasks: the greater the information content of the subsumer, $d = -\log p(sub(t_1,t_2))$, the more similar the tasks $t_1$ and $t_2$ are.
\end{itemize}
Consequently, Li et al.\ define the first measure to be anti-monotonic in the range $[0,1]$ and the other measures to be monotonic in the same range, where $0$ represents complete dissimilarity, and $1$ represents complete similarity. The contribution of each of these measures is taken independently of each other:
\begin{equation}\label{eq:sim}
sim(t_1,t_2) = e^{-\alpha l} \cdot \tanh \beta h \cdot \tanh \lambda d
\end{equation}
Constants $\alpha$, $\beta$ and $\lambda$ are positive real numbers that determine the relative influence of each of the three measures on the final similarity. They provide us with suitable adjustment points for the overall semantic similarity to fit well and evolve with the actual usage of terms by a concrete community.

%For the semantic similartiy measure used by the trust model, we combine the propagation of an evaluation of the activity performed in one task to the activity in the other (according to the activity meronomy) with the value for the similarity of the activities' objects (according to the object taxonomy). % NAR XXX this is not true, the combination is done at the trust level at the moment.

\subsection{Trust model}\label{sec:trust}
As illustrated earlier, the flooding algorithm is designed to flood to trustworthy nodes. What is important then is to calculate the requester's trust, with respect to a given request, on the reachable nodes of the network in order to assess whether these nodes can be asked for help or not. This section presents the trust model that calculates these required trust measures. 

We say every member maintains his own trust evaluations of his friends (his neighbouring nodes in the network). These trust evaluations are task-dependent, with which we mean that a user's evaluation of another changes with respect to the task for which he is evaluated. %This trust evaluation is crucial in the functioning of the flooding algorithm, which only forwards a request for help to a neighbouring node if the trust level on that node is sufficiently high. 
%In other words, w
We assume that these evaluations have a transitive property: that is, if we know $a$'s trust on $b$ ($trust(a,b)$) and $b$'s trust on $c$ ($trust(b,c)$) then we can calculate $a$'s trust on $c$ ($trust(a,c)$). It is this transitivity that allows requests to propagate in a network. 
The transitivity of trust is not uncommon in the trust literature. For instance, TidalTrust~\cite{Golbeck:2006:CPT:2165554.2165569} is a trust model that propagates trust evaluations through a social network, although a more sophisticated algorithm is used to compute the trustworthiness of a node. For a more in-depth discussion on transitivity of trust and considerations that should be taken into account when propagating trust through a network, we refer the interested reader to~\cite{Josang:2006:SAT:1239776.1239778,FalconeCastelfranchi10}. 
In our case, and as illustrated earlier, we assume trust is monotonically decreasing along a single path. This is modelled using a T-norm, and we use the minimum ($\min$) as the T-norm function. 

The trust mechanism implemented in \uhelp\ makes use of the ratings that requesters provide when a volunteer completes his task. A rating $E$ is a tuple of the form:
$$\langle Requester, \mathit{Volunteer}, Activity, Object, \mathit{Value} \rangle$$
where $Requester$ is the person that asked for help, $\mathit{Volunteer}$ is the volunteer chosen to carry out the task, $Activity$ is the task for which help is required (it is an element of the meronomy of activities, as illustrated in Section~\ref{sec:semantics}), $Object$ is the object associated with the activity (such as the relative when caring for a relative or the item when lending something, and it is an element of the taxonomy of objects, as illustrated in Section~\ref{sec:semantics}), and $\mathit{Value} \in [1,7]$ with 1 representing complete failure and 7 representing complete satisfaction.\footnote{We choose the scale $[1,7]$ based on Miller's Law in Psychology, which argues that the number of objects than an average human can hold in his memory is $7 \pm 2$~\cite{seven}. Other scales can easily be applied here.}  We refer to the elements in the tuple using superscripts. For instance, to refer to the $Activity$ we use the notation $E^{Activity}$. And we refer to the set of evaluations of a user $X$ as $\mathcal{E}_{X}$.

% NAR XXX: commented what is below for now
%If the activity includes other sub-activities, the requester can decide to evaluate the general activity only, or to perform a detailed evaluation of each one of the leaf activities. For instance, if the activity was babysitting, then the user can also evaluate changing nappies and/or feedings, and so on.\footnote{The current implementation of \uhelp\ does not permit evaluation of sub-activities, since these are not explicitly defined.} % NAR XXX: why is the following important.
% The ChildType is always a leaf in the Children's taxonomy.

The basic idea behind the trust model is to learn from \emph{similar} past experiences. For instance, if one is great at donating stuff (that is, previous ratings on his donation activities have been high), then he might be considered good at lending stuff as well, since donating and lending are similar activities.  In this manner, we learn one's trust on another with respect to a given activity by learning from similar past experiences. However, we do not \emph{only} consider the similarity of activities, but the activities' objects as well. For instance, one may be a great babysitter, but terrible at caring for an elderly. As such, we also consider the similarity of the activities' objects.

%Some examples of evaluations:
%\begin{itemize}
%\item $(Ann123, John243, ``entertaining", ``toddler", 0.7)$
%\item $(Paul324, Mary334, ``singing songs", ``preschooler", 0.8)$
%\end{itemize}

We then calculate the trust of a person $R$ on some potential volunteer $V$ at time $t$ with respect to an activity $A$ and its object $O$ as a function that aggregates $R$'s trust on $V$ with respect to activity $A$ and $R$'s trust on $V$ with respect to object $O$:

\begin{equation}\label{eq:trust}\small
\mathit{Trust}^{t}(R,V,\langle A,O\rangle) = \alpha \cdot \mathit{Trust}_{OBJ}^{t}(R,V,O) + (1-\alpha) \cdot \mathit{Trust}_{ACT}^{t}(R,V,A)
\end{equation}
where $\alpha$ is the weight given to each type of trust: the trust with respect to an activity and that with respect to an object, which we discuss next in more detail.

\subsubsection{Trust with respect to an activity}
We first recall that activities are defined through meronomies by the part–of relation. For instance, changing nappies and feeding a baby are parts of the babysitting activity (Figure~\ref{fig:activity-meronomy}).  As such, to evaluate the trust of an individual regarding the activity that she is requested to perform ($\mathit{Trust}_{ACT}^{t}(R,V,A)$), we make use of the OpinioNet algorithm~\cite{Osman2010}. This is because OpinioNet is an algorithm that propagates opinions (ratings, in our case) from one node to another in a meronomy. For instance, if one received ratings on her babysitting abilities, then we can assess her changing nappies and baby feeding abilities. And vice versa. 

%%%%%%%%%%%
The basic idea behind OpinioNet is that if a node in the meronomy does not receive a direct evaluation (or a rating), then its evaluation may be deduced from its children nodes' evaluations. This is because the parent node is structurally composed of its children nodes. Hence, the evaluations on children nodes must necessarily influence the deduced evaluation on a parent node. OpinioNet refers to the direct evaluation on a node or an evaluation of it that is deduced from evaluations of the parts that compose it as the `intrinsic evaluation' of that node.

Additionally, an `extrinsic evaluation' is an evaluation that is propagated down from parent nodes to children. In the absence of information about the node itself, or the parts that compose it, information may be inherited from what one belongs to. In other words, in the absence of information about intrinsic evaluations, the evaluation of that node is calculated based on evaluations of its parents' nodes.

As an example of how the OpinioNet algorithm propagates evaluations through the meronomy, consider that if a potential volunteer performed well at the \emph{preparing meal} activity, then this will affect his evaluation for the \emph{babysitting} activity. This is considered an `intrinsic evaluation' of the \emph{babysitting} activity. An `extrinsic evaluation' (of the \emph{changing nappies} activity) would be to say that performing well at \emph{babysitting} carries over to a good evaluation for the \emph{changing nappies} subactivity, when there is no direct evaluation of the potential volunteer with respect to \emph{changing nappies}.
%%%%%%%%%%%

In the case of \uhelp, we say for a given meronomy $M$ ($M=\langle \mathcal{O}, \mathcal{P} \rangle$, where $\mathcal{O}$ is the set of objects and $\mathcal{P}$ is the set of edges, or part of relations between objects) and a set of ratings provided by $R$ on $V$ with respect to different activities in the meronomy ($\mathcal{R}_{(R,V)} : \mathcal{O} \to \mathit{Value}$), OpinioNet is capable of calculating the trust on the remaining activities in the meronomy. To calculate the trust on a specific activity $A\in M$, we then have:
\begin{equation}\label{eq:trust_act}
\mathit{Trust}_{ACT}^{t}(R,V,A) = OpinioNet(A,M,\mathcal{R}_{(R,V)})
\end{equation}

For further details on OpinioNet, we refer the interested reader to \cite{Osman2010}.  

\subsubsection{Trust with respect to an object}
Objects are usually defined through taxonomies by the is-a relation. Take the taxonomy of relatives for instance, we say a `newborn' is an `infant', and an `infant' is a `kid', and so on (Figure~\ref{fig:child-taxonomy}). 

Our evaluation of trust with respect to an object depends on learning from past experiences (whose outcome is specified through ratings) of similar objects. The similarity between two objects ($sim(O_i, O_j)$) was presented by Equation~\ref{eq:sim}, and it is based on the location of the objects in the taxonomy of objects. However, we use a similarity threshold $\zeta$ to filter out the evaluations associated with distant (non-similar) objects. In other words, we modify the semantic similarity of objects defined by Equation~\ref{eq:sim} accordingly:
\begin{equation}
sim_{OBJ}(O_i,O_j) = \left\{
	\begin{array}{l l}
	sim(O_i, O_j) & \mathit{if} sim(O_i, O_j) > \zeta\\
	0 & otherwise
	\end{array} \right.
\end{equation}
where $sim(O_i,O_j) \in [0,1]$ is the semantic similarity between objects $O_i$ and $O_j$ (as defined by Equation~\ref{eq:sim}). Below the threshold $\zeta$, the similarity between two objects is considered too low to make use of the associated experience, and as such, we re-set this similarity to 0 so that it is disregarded in the trust calculation.

The trust that requester $R$ gives to volunteer $V$ with respect to a particular object $O$ at time $t$ is then calculated by following a weighted aggregation of all the ratings of past experiences, where the weights are the similarities of those experiences with respect to their objects:
\begin{equation}\label{eq:trust_obj}
\mathit{Trust}_{OBJ}^{t}(R,V,O)= \frac{\displaystyle\sum_{E_i \in \mathcal{E}_{R}(V)^{t}} (sim_{OBJ}(O, E_i^{Object}) \cdot E_i^{\mathit{Value}})}{\displaystyle\sum_{E_i \in \mathcal{E}_{R}(V)^{t}} sim_{OBJ}(O, E_i^{Object})}
\end{equation}
where $\mathcal{E}_{R}(V)^{t}$ is the set of all evaluations made by $R$ on $V$ by time $t$.
% NAR XXX why are we using a window?!
% time window $[t-TWin,t]$, where $TWin$ is a predefined parameter.

\subsubsection{Extending the trust model with shared ratings}
In the trust model proposed above, we assume each user maintains its own ratings on others, and makes use only of these ratings when assessing others' trustworthiness. That is, it relies on its \emph{personal} past experiences only. Formally, this is specified by conditioning the set of evaluations maintained by a user $X$ ($\mathcal{E}_{X}$) to contain ratings provided by $X$ only: 
$\forall E_{i} \in \mathcal{E}_{X} \cdot E_{i} = \langle R,V,A,O,T\rangle \wedge R=X$. 

In this section, we propose an extension to the previous model by allowing users in a social network to share their ratings. That is, the set of evaluations that a user has access to may contain evaluations from other users in the social network. As a result, calculating the trust on someone with respect to a given task is then based on the (accessible) collective ratings on this person for similar past experiences.%\footnote{We distinguish between trust and reputation, and we say trust represents one's XXX of another while reputation represents XXX. Stilstill redfer to this measure as trust as opposed to reputation since the measure is. XXX}

To achieve our goal, first a model for sharing ratings should be selected, which is a sensitive issue that usually raises privacy concerns. A safe option that we recommend is to have the user manually select, from her contact list, the friends that she accepts to share ratings with. She may also select the friendship level for the propagation of ratings. For instance, one may specify that they are willing to share their ratings with trusted friends of their trusted friends. An alternative method to manual selection, although less secure, is to allow one to share her ratings with another when their ratings are similar. In other words if two people hold similar opinions about others performances, then they may share these opinions. The similarity of ratings is specified as $w(R,U,X)$ (which specifies the similarity of the ratings of $R$ and $U$ with respect to an issue $X$), and they are discussed shortly. Yet another alternative is to allow one to share her ratings on a task $\langle A,O\rangle$ (that is, a task with activity $A$ and its object $O$) with those that it trusts sufficiently enough with respect to performing that task $\langle A,O\rangle$: that is, a person $R$ shares its rating on a task $\langle A,O\rangle$ with another person $V$ at time $t$ if $\mathit{Trust}^{t}(R,V,\langle A,O\rangle) > \eta$, where $\eta$ specifies the threshold that deems a person sufficiently trustworthy to share ratings with. These sample alternative methods for sharing ratings will require further exploration and evaluation with real user communities.

Given that ratings are shared (that is, $\mathcal{E}_{R}$ contains both ratings made by $R$ and those shared with $R$), trust is then modified from Equation~\ref{eq:trust} accordingly:
\begin{equation}\label{eq:trustextended}\small
\begin{array}{lll}
\mathit{Trust}^{t}(R,V,\langle A,O\rangle) & = &
    \alpha \cdot \frac{\displaystyle\sum_{\forall U \in raters(\mathcal{E}_{R}^{t}(V))} w(R,U,O) \cdot \mathit{Trust}_{OBJ}^{t}(U,V,O)}{\displaystyle\sum_{\forall U \in raters(\mathcal{E}_{R}^{t}(V))} w(R,U,O)} \\
    & + & (1-\alpha) \cdot \frac{\displaystyle\sum_{\forall U \in raters(\mathcal{E}_{R}^{t}(V))} w(R,U,A) \cdot \mathit{Trust}_{ACT}^{t}(R,V,A)}{\displaystyle\sum_{\forall U \in raters(\mathcal{E}_{R}^{t}(V)) w(R,U,A)}}
\end{array}
\end{equation}
where $\alpha$ is the weight given to each type of trust, $\mathcal{E}_{R}^{t}(V)$ is the set of all ratings (made by or shared with $R$) that assessed $V$ by the time $t$, and $raters(\mathcal{E}_{R}^{t}(V)) = \{ U \,|\, \langle U,V,X,Y,Z\rangle \in \mathcal{E}_{R}^{t}\}$ is the set of requesters that have rated the volunteer $V$ and shared their ratings with $R$ (by time $t$).

In summary, instead of combining the trust of requester $R$ on potential volunteer $V$ with respect to the object $O$ and activity $A$ based on a pre-defined weight $\alpha$ (Equation~\ref{eq:trust}), we combine the \emph{trust of a group of raters} (who have shared their ratings with $R$) on potential volunteer $V$ with respect to $O$ and $A$ in a similar manner (Equation~\ref{eq:trustextended}). The trust of a group of raters on $V$ with respect to $O$ (resp. $A$) is calculated by aggregating the individual trust of each rater in the group on $V$ with respect to $O$ (resp. $A$). The aggregation is a weighted aggregation, where the weight given to the trust of each individual rater $U$ is $w(R,U,O)$ (resp. $w(R,U,A)$). This weight represents the trust that the requester $R$ holds on the rater $U$ with respect to the object $O$ (resp. activity $A$). This trust measure, $w(R,U,O)$ (resp. $w(R,U,A)$), is different than that of Equation~\ref{eq:trust_obj} (resp. Equation~\ref{eq:trust_act}), in the sense that this measure describes how much is the user $U$ trusted with respect to \emph{giving a rating} concerning object $O$ (resp. activity $A$), as opposed to how much is he trusted with respect to performing a task whose object is $O$ (resp. activity $A$). 

We say trust on giving ratings may be learnt from the similarity of past ratings. For this, we make use of the COMAS model~\cite{comas}. In summary, COMAS tries to learn how much trusted are two opinion holders (or raters) based on the similarity of their ratings. It is based on two intuitions. First, if two users have both rated the same object $O$ (activity $A$) in the past, then the similarity of their ratings can be used to describe the trust they have on each others' future ratings. That is, the more similar their rating profile, the more trusted are the users with respect to each other.  This is referred to as the direct trust between two users. When there are no common objects (activities) rated by both users, COMAS approximates the unknown trust between them through the transitivity of trust over the path with direct trust that links those two peers in question. And this is referred to as the indirect trust between two users. For further details on calculating $w(R,U,O)$ and $w(R,U,A)$, we refer the interested reader to~\cite{comas,paas}.\footnote{While PAAS~\cite{paas} is another model that achieves the same objectives as COMAS~\cite{comas}, even following the same intuition as COMAS, PAAS is a bit more sophisticated as it uses probability distributions to specify ratings and their similarity. This makes PAAS a richer and more informative model as much more information is preserved in the calculations, allowing PAAS to compute the uncertainty of group opinions.} % NAR XXX ?! is this true?! COMAS and PAAS may be used interchangeably in our case.

\section{Implementation}\label{sec:implementation}
This section presents the details of implementing the \uhelp\ app. %The implementation was carried out using the Ionic framework, \url{ionicframework.com}, which allows developing an application as a web application that would additionally automatically be translated into iOS and Android applications. This allowed us to build a cross-platform application. We also made use of the Apache Cordova platform, \url{cordova.apache.org}, for developing plugins that map device hardware to plugins in javascript, which were then used on the Ionic framework. This was basically needed to access necessary device features, such as contacts, notifications, etc., using a javascript call. 
%
%In order to use \uhelp\ it is necessary for the user to configure her data upon downloading the application.  She must first either login via Facebook or register with the system and import her connections from her list of contacts, to create her social network.  Once this is done, the user can start to use \uhelp\ to request help, and be contacted by others in the community who need help in return.
%To use \uhelp, however, we note that one should first register with \uhelp\. This is done when the app is first downloaded. The welcome screen will allow one to sign in (if they already have an account with \uhelp), register by providing a phone number and a password, or login via Facebook. After logging in, the app will ask the user to access her contacts list, which is crucial for building one's social network, and enable notifications. The user is then considered setup to use \uhelp.
%
In what follows, we present the user interface (Section~\ref{sec:interface}), the functionality behind that interface (Section~\ref{sec:fsm}), and some notes on the \uhelp\ architecture (Section~\ref{sec:architecture}).

\subsection{User Interface}\label{sec:interface}
In this section we present the user interface of the \uhelp\ application. The \uhelp\ application's usability is designed around four main views: (1) \emph{the ``Help'' view}, where users may request help; (2) \emph{the ``Requests'' view}, where users can track and manage requests for help (issued by them or others); (3) \emph{the ``Community'' view}, where users can track and manage their community, or social network; and \emph{the ``Settings'' view}, where users can change their user-specific settings or get more information about the app. 

To use \uhelp, however, we note that one should first register with \uhelp. This is done when the app is first downloaded. The welcome screen will allow one to sign in (if they already have an account with \uhelp), register by providing a phone number and a password, or login via Facebook. After logging in, the app will ask the user to access her contacts list, which is crucial for building one's social network, and enable notifications. The user is then considered setup to use \uhelp.

Next, we go over each of the four main \uhelp\ application views in detail.

\subsubsection{Asking for help}\label{sec:helpview}
\begin{figure}[!ht]
  \centering
  \begin{minipage}[t]{0.26\textwidth}
    \includegraphics[width=\textwidth]{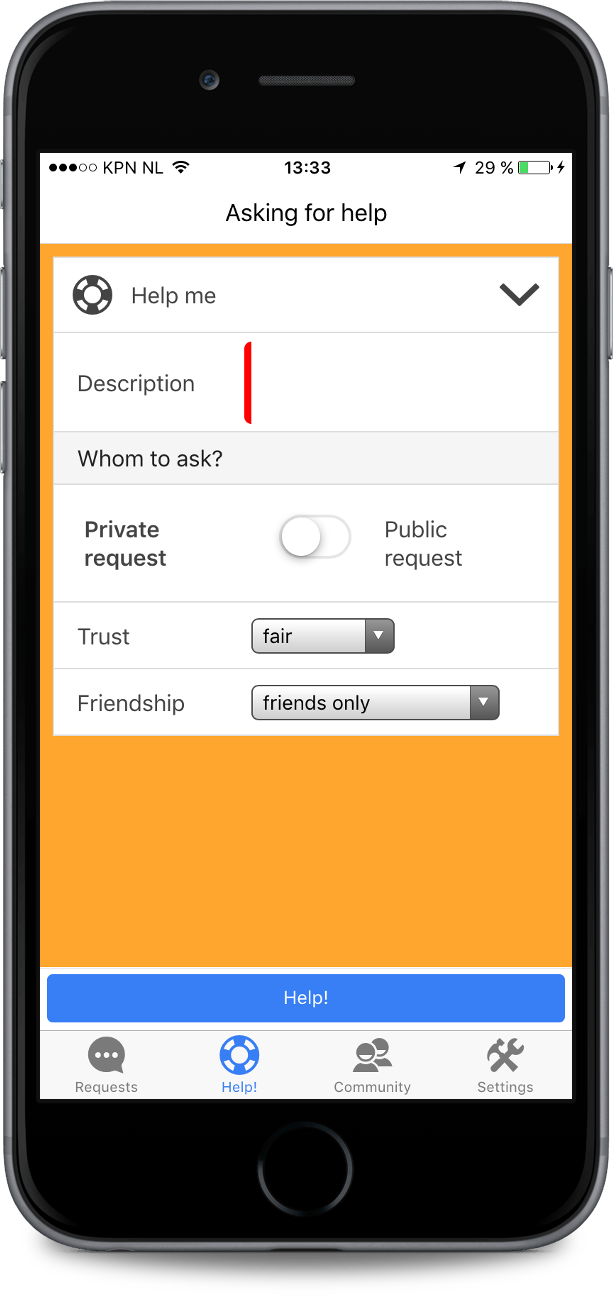}
    \caption{Help view, for general requests}
    \label{fig:helpgeneral}
  \end{minipage}
  \hfill
  \begin{minipage}[t]{0.26\textwidth}
    \includegraphics[width=\textwidth]{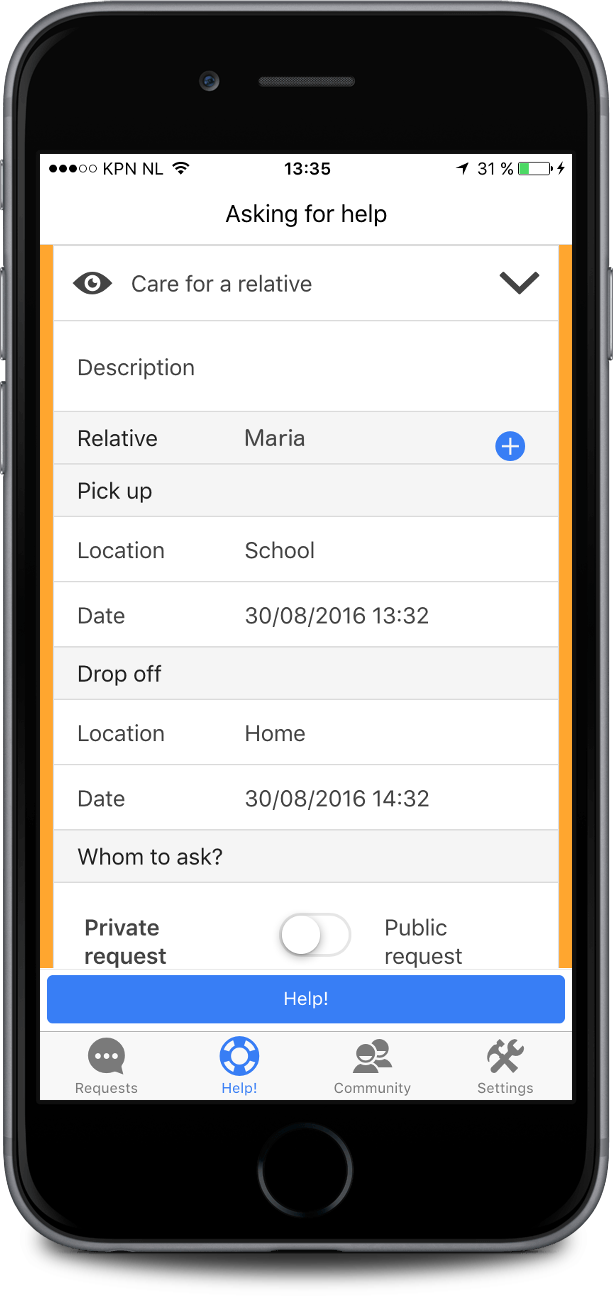}
    \caption{Help view, for caring for a relative}
    \label{fig:helpchild}
  \end{minipage}
  \hfill
  \begin{minipage}[t]{0.26\textwidth}
    \includegraphics[width=\textwidth]{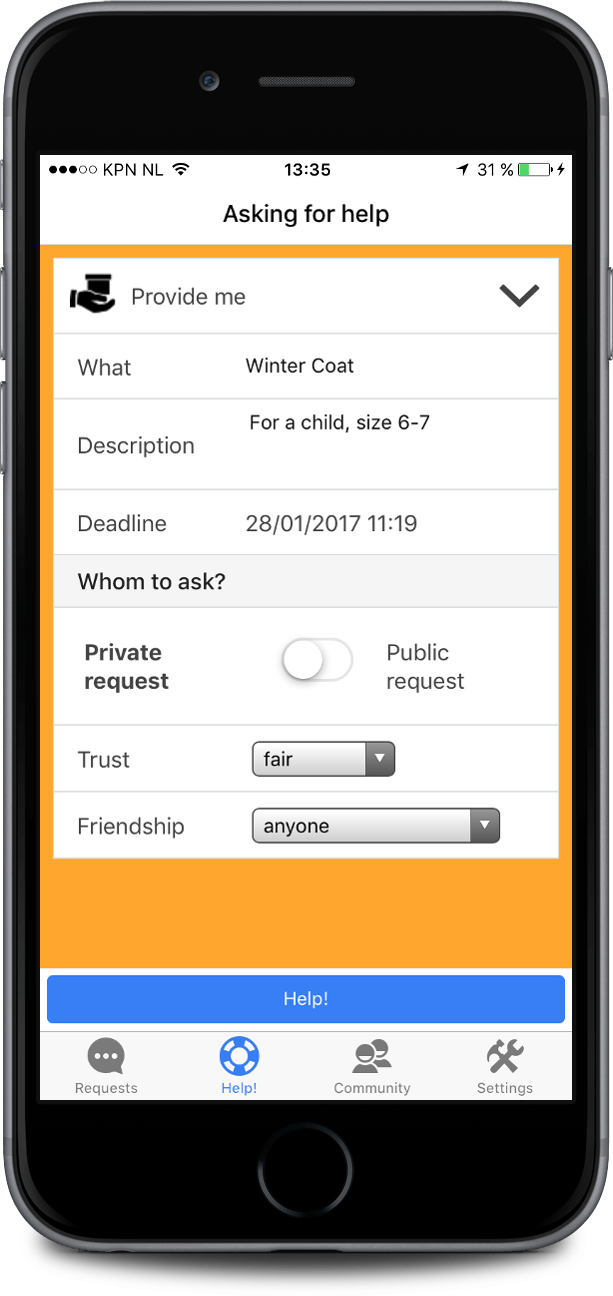}
    \caption{Help view, for donating}
    \label{fig:helpprovide}
  \end{minipage}
\end{figure}

\begin{figure}[!hb]
  \centering
  \begin{minipage}[t]{0.26\textwidth}
    \includegraphics[width=\textwidth]{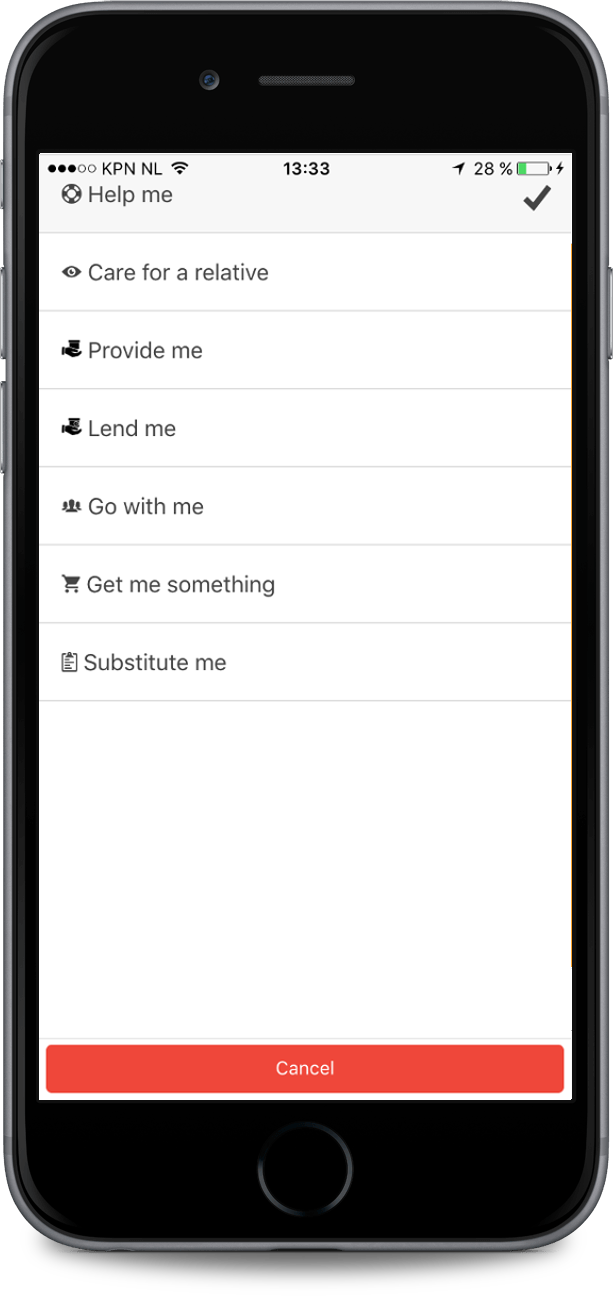}
    \caption{Task types}
    \label{fig:tasks}
  \end{minipage}
  \hfill
  \begin{minipage}[t]{0.26\textwidth}
    \includegraphics[width=\textwidth]{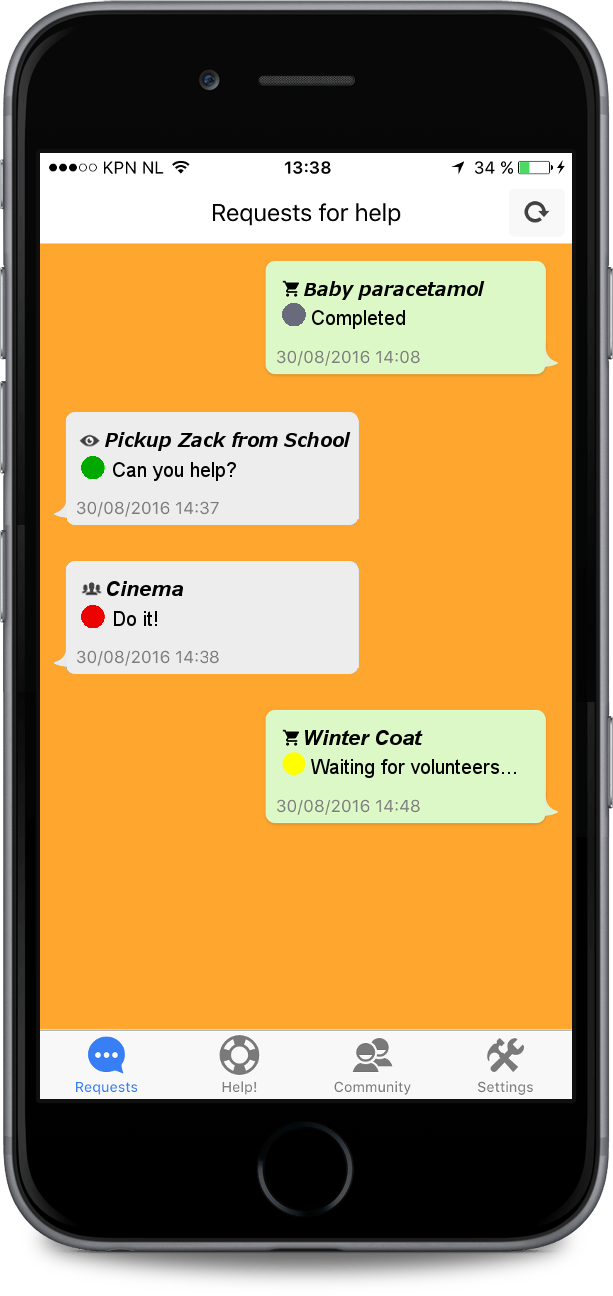}
    \caption{Requests}
    \label{fig:requests}
  \end{minipage}
  \hfill
  \begin{minipage}[t]{0.26\textwidth}
    \includegraphics[width=\textwidth]{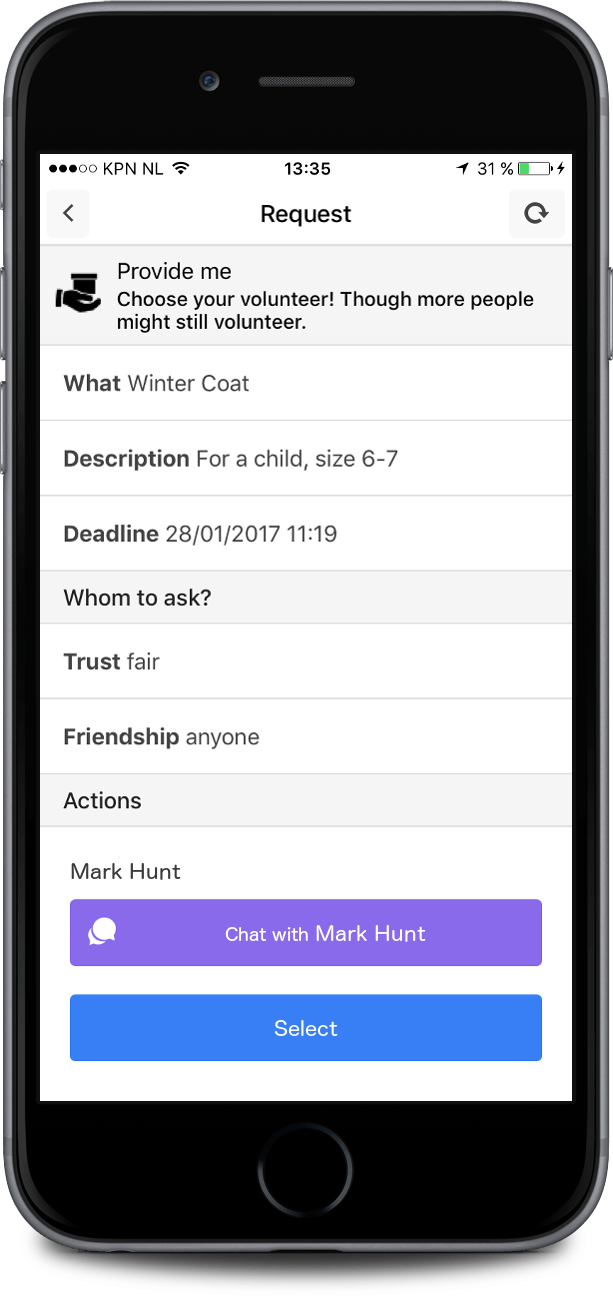}
    \caption{Request details}
    \label{fig:request}
  \end{minipage}
\end{figure}

Anyone can ask for help at anytime. However, when asking for help (the ``Help'' view, Figure~\ref{fig:helpgeneral}), the user should first choose what type of task does he need help with (Figure~\ref{fig:tasks}). There are seven different tasks in total. The first is a general task, entitled ``Help me", where the user can specify their request by filling a single field, the description field. The remaining six tasks are more specific, and they have been designed by taking into consideration the requirements of our use case community, a community of single parents in Barcelona, which we discuss in more detail in Section~\ref{sec:usecase}. These tasks are (1) ``Care for a relative", which allows one to find volunteers to care for a relative, like an elderly or a child, including taking that relative from one place to another, and the requester will need to specify the relative to be taken care of along with the pickup and drop off locations and dates/times; (2) ``Provide me", which allows one to ask for donations, and the requester will need to specify the item that he requires and the date/time of when is this item needed; (3) ``Lend me'', which allows one to ask to borrow something, and the requester will need to specify the item that he needs to borrow and the start and end date/time for borrowing this item; (4) ``Go with me'', which allows one to look for company for some activity that he plans to do, such as go to the park or the theatres, and the requester will need to specify the activity that he would like company with, and the location and date/time of the activity; (5) ``Get me something'', which allows one to ask for something to be delivered, say if one is ill in bed and they require someone to get them the medication or an elderly requiring someone to get him the groceries, and the requester will need to specify the item(s) they need, and the location and date/time for dropping off the item(s), although a pickup location and date/time is also possible to specify but is left optional; and finally (6) ``Substitute me'', which allows one to find substitutes, say for work, and the requester will need to fill in the activity that they are looking for substitutes for and its start and end date/time. 

After specifying the task, the user will need to specify \emph{whom} to ask for help. Here, there are two main options, make the request (1) a public request, which implies that any person who has \uhelp\ installed will receive this request; and (2) a private request, which implies that only one's social network will be crawled. In the latter case, the user may further specify the trust and friendship levels. In other words, they specify the minimum level of trust acceptable for this specific task, and the maximum number of hops allowed when crawling their social network. For example, for picking up one's son from school, a mother may require \emph{only} friends with utmost trustworthiness to be asked. For a similar request, another parent may accept friends of friends. However, to get the medication delivered, an elderly person may loosen the trust and friendship requirements further. 

The entire process of specifying what is the help needed and whom to ask for help is in fact achieved with just a few clicks, as illustrated by Figures~\ref{fig:helpgeneral}--\ref{fig:helpprovide}. Only after filling in the required parameters can the user proceed to pressing the ``Help!'' button. This button then triggers the flooding algorithm (Section~\ref{sec:flooding}) that spreads the help request appropriately in order to find suitable volunteers satisfying the requester's requirements. The task is automatically added to the requester's list of requests. Furthermore, every time the flooding algorithm asks a community member for help, a notification is displayed on that person's mobile or tablet, and the request is added to their list of requests.

%However, the user must first choose the action that she requires help with from the list tasks.  The elements of this list, and the options given, correspond to elements in the predefined ontology.  We note that different actions will require different parameters.  To keep the application user friendly, once an action is chosen from the list (Figure~\ref{fig:uhelp-allhelp}), the appropriate fields will appear, where the user can fill in the details of the requested action (Figure~\ref{fig:uhelp-pickuphelp}). For example, if the action is ``Care for a relative", then the relative field appears, along with the pickup and drop-off location and time, but if the action is "Get me something", then the object field appears, along with the pickup and drop-off location and time.

\subsubsection{Managing requests}\label{sec:requestsview} 
Users can track and manage the requests for help (sent by or to them) in the ``Requests'' view, Figure~\ref{fig:requests}. Requests are represented by text bubbles. Green text bubbles (to the right) represent one's own requests, while grey ones (to the left) represent requests received. Each text bubble shows (1) the type and name of the request, along with who made that request (in the case of requests received), (2) the state of the request, (3) the deadline for the request, or the date/time by which the request should be completed; and (4) a notification count which pops up when there are new chat messages. See Figure~\ref{fig:requests} for illustration.

%Again, the presentation of each item in the list is dependant on the semantics of the action.  In the case of the ``Care for a relative'' action, the name of the child and the pickup location will be displayed.  Additionally, the ``state'' parameter is presented to help the user track the progress of requests, such as `waiting for volunteers', `help on its way', or `completed'.

\begin{table}[!b]\footnotesize
\centering
\begin{tabular}{p{3.5cm}p{1cm}p{7.5cm}}
{\bf State} & {\bf Colour} & {\bf Description} \\ \hline
looking for volunteers & yellow  & the request awaits volunteers \\ 
pending assignement 1 & green  &  volunteer(s) found, more may be on their way \\ 
pending assignement 2 & red  &  the request awaits the selection of a volunteer \\ 
assigned & yellow &   the request was assigned to a volunteer \\ 
completed & green &   the request was completed and awaits being rated \\ 
rated & grey & the request was completed and rated  \\ 
cancelled & grey & the request was cancelled  \\ 
expired & grey & the request expired  \\ \hline
\end{tabular}
\caption{The states of a requester's request}\label{tbl:requester}
\end{table}
\begin{table}[!b]\footnotesize
\centering
\begin{tabular}{p{3.5cm}p{1cm}p{7.5cm}}
{\bf State} & {\bf Colour}  & {\bf Description} \\ \hline
unanswered & green & the requestee has not responded yet \\ 
declined & grey & the requestee declined from volunteering \\ 
accepted & yellow & the requestee volunteered, waiting to be chosen \\ 
committed & red &  the requestee has been chosen \\ 
help not needed & grey & the requestee has not been chosen \\ 
completed & grey & the request was completed \\ 
cancelled & grey & the request was cancelled \\ 
expired & grey & the request expired \\\hline
\end{tabular}
\caption{The states of a requestee's request}\label{tbl:requestee}
\end{table}

We note that the state for the requestee is different from that of the requester. Tables~\ref{tbl:requester} and~\ref{tbl:requestee} present the different states for each, along with how is this state visualised to the user. The visualisation is executed by assigning colours to states (Figure~\ref{fig:requests}). We use colours to categorise states into four categories: (1) those that need to be acted upon, which get the colour green; (2) those that need to be acted upon \emph{immediately} (for example, when ones needs to urgently select a volunteer as the deadline to select volunteers is approaching), which get the colour red to indicate urgency; (3) those that require no action at the moment as others are expected to be acting upon those requests, which get the colour yellow; and (4) those that are considered closed, which get the colour grey. The states are discussed in further detail in Section~\ref{sec:fsm}.

When selecting a specific request, the details of that request are presented (Figure~\ref{fig:request}). It is in this view where a requester/requestee can manage a request. We note that the set of actions that a requester can take here are different from those of the requestee.  
The requester's actions that are permitted in this view are: (1) \emph{cancelling a request}, which is allowed at any time from when the request is made up until the request is completed; (2) \emph{contacting volunteers}, where the requester can have a private chat or even phone call with the volunteers if needed (and if the privacy settings permit this, as we illustrate in Section~\ref{sec:settingsView}), and this is allowed at any point when there are volunteers and up until the request is completed; (3) \emph{selecting a volunteer}, where the requester can select her volunteer from the list of available volunteers, if any, though there is a deadline for selecting volunteers as illustrated by Section~\ref{sec:fsm}; and  (4) \emph{rating a volunteer}, where the requester may rate a volunteer only allowed if the request has been marked as completed by the selected volunteer or the request's deadline has passed. 

The requestee's actions that are permitted in this view are: (1) \emph{volunteering}, where the requestee accepts to volunteer, and this is allowed from the moment a request is received up until a deadline for volunteering is reached; (2) \emph{declining from volunteering}, where the volunteer declines from volunteering, and this is also allowed from the moment a request is received up until a deadline for volunteering is reached; (3) \emph{cancelling}, where a requestee can cancel after accepting to volunteer, and the cancel button should be active from the moment the requestee accepts a request up until the request's deadline; (4) \emph{contacting the requester}, where the requestee can have a private chat or even phone call with the requester if needed, and this is allowed from when the requestee volunteers up until the request is completed; and (5) \emph{completing a request}, where the requestee can mark a request as completed, which is only allowed after the requestee has been selected for this request and up until the request's deadline. Note that as mentioned above there is a deadline for volunteering, and up until this deadline is reached, the requestee may change its mind over and over again, allowing her to move between accepting and cancelling as many times as she so wishes.

% NAR XXX As Section~\ref{fsm} illustrates, there are different final states to a request, such as expired, cancelled, and completed. All of the above actions are naturally allowed in non-final states only.

\subsubsection{Managing one's social network}\label{sec:communityView}
\begin{figure}[!ht]
  \centering
  \begin{minipage}[t]{0.26\textwidth}
    \includegraphics[width=\textwidth]{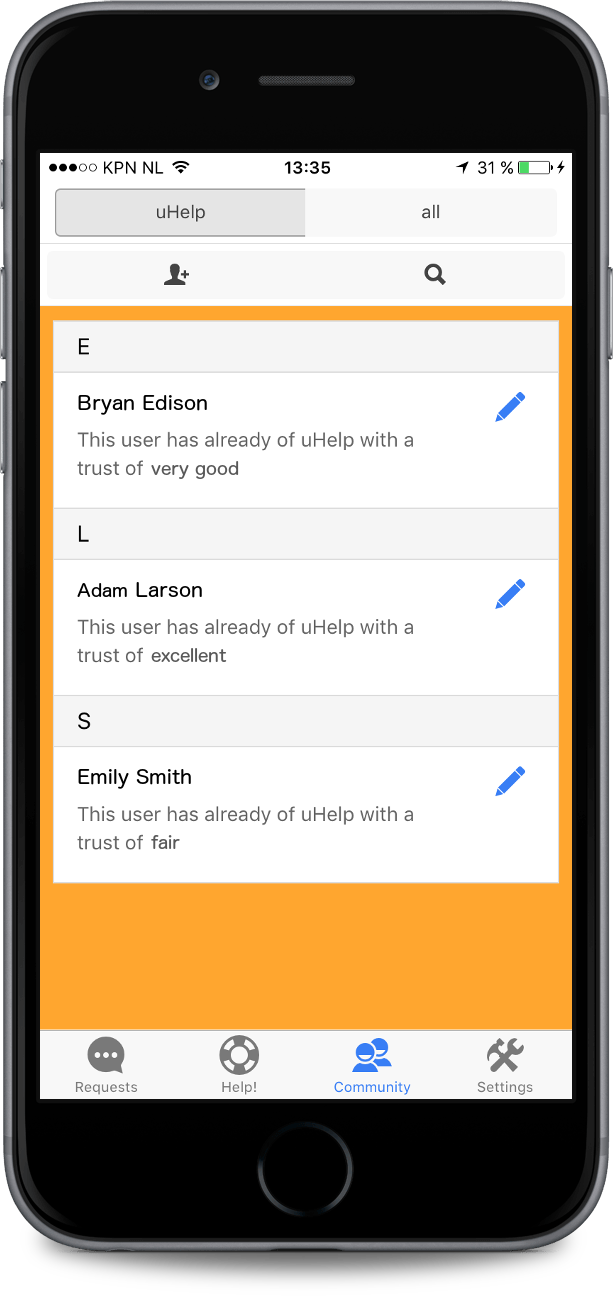}
    \caption{Community}
    \label{fig:community}
  \end{minipage}
  \hfill
  \begin{minipage}[t]{0.26\textwidth}
    \includegraphics[width=\textwidth]{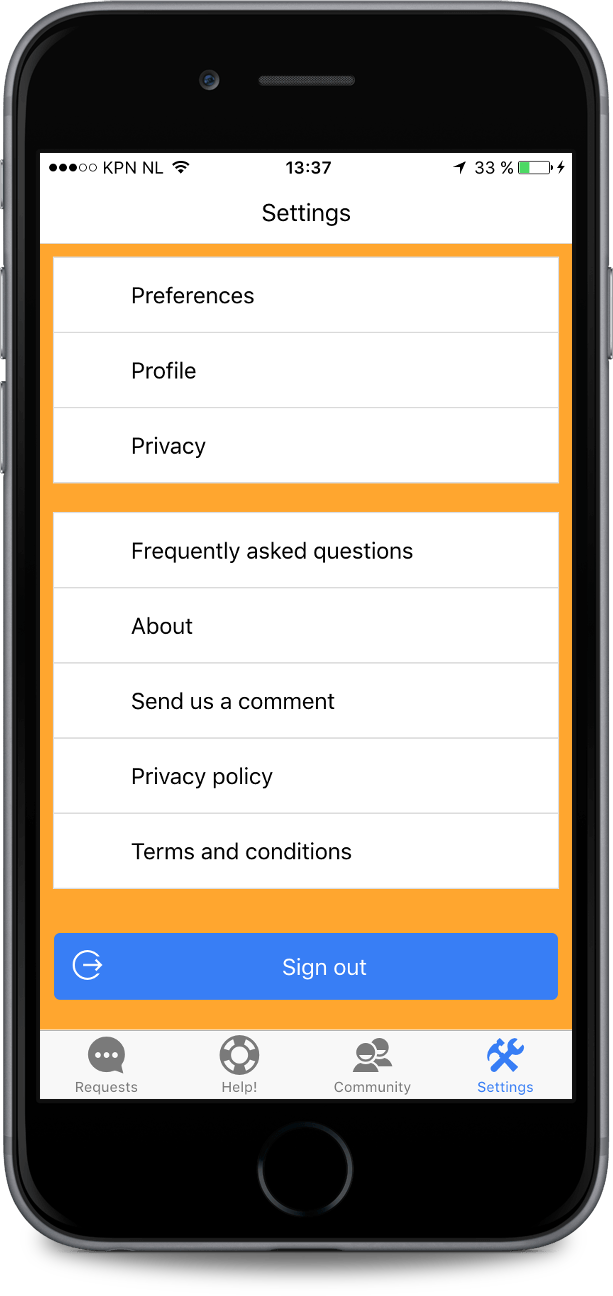}
    \caption{Settings}
    \label{fig:settings}
  \end{minipage}
  \hfill
  \begin{minipage}[t]{0.26\textwidth}
    \includegraphics[width=\textwidth]{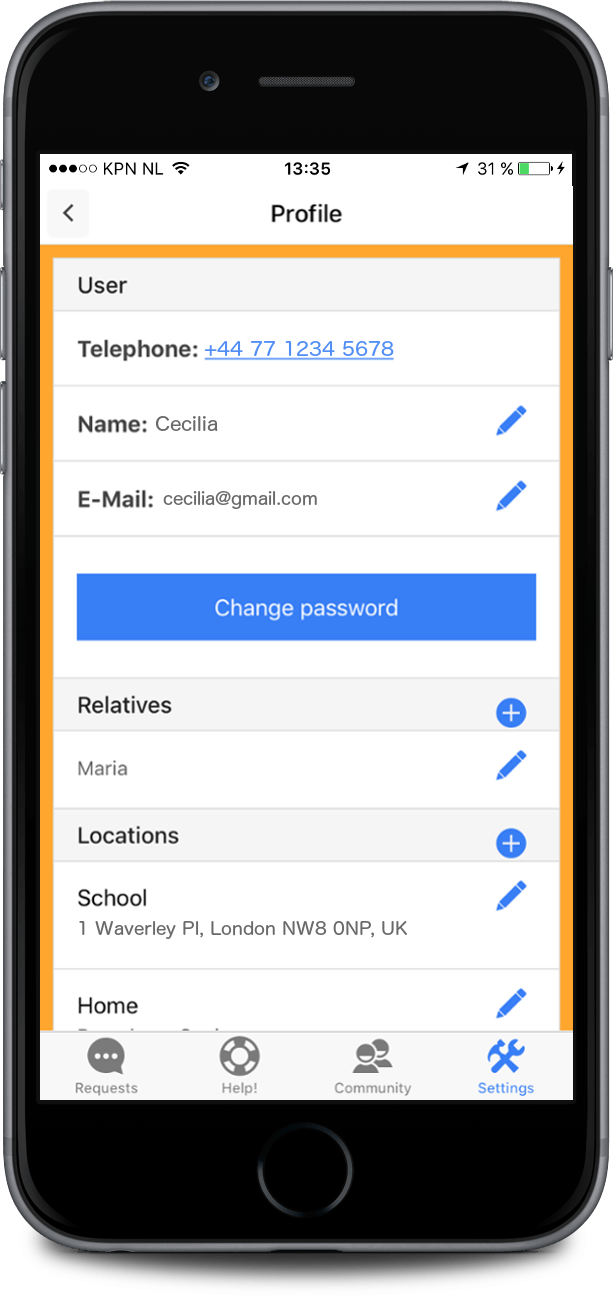}
    \caption{Profile}
    \label{fig:profile}
  \end{minipage}
\end{figure}

In the ``Community'' view, Figure~\ref{fig:community}, a user can see the list of ``all'' his contacts (this list is initially imported from the user's mobile phone or tablet when the app is first installed, and then updated every time the app is launched), or a list of those contacts who have \uhelp\ installed (this list is also updated every time the app is launched). When in the ``all'' contacts view, contacts who do not have \uhelp\ installed may be invited to install \uhelp. This is done by sending them an email or whatsapp message with a link to the app.  Contacts who do have \uhelp\ installed are those that \uhelp\ considers as friends, and they are one hop away in one's social network in \uhelp. The trust that one holds on his friends may be updated manually in the ``Community'' view. The basic idea is that one's trust on any friend starts with an average level (specified as `fair'). Then, every time this person volunteers they are rated for their performance. The trust model of Section~\ref{sec:trust} illustrated how this trust level gets updated with every rated performance. Nevertheless, at any point in time, one can update these numbers manually in the ``Community'' view. Recall that one's trust level is calculated with respect to a given community member. In other words, while John may not be pleased with Anna's performance, Adam may be more than happy. As such, John's trust on Anna may be low, whereas Adam's trust on Anna may be high.

\subsubsection{Settings}\label{sec:settingsView}
The ``Settings'' view, Figure~\ref{fig:settings} has a number of sub-views. Some of them are informative, such as accessing the frequently asked questions, the `About' section, the privacy policy, the terms and conditions of usage, as well as sending feedback to the developers. 

Other sub-views allow the user to edit their preferences, profile, and privacy settings. For instance, under the ``Preferences'' sub-view, the user may select the language of the app. We currently have \uhelp\ in English, Spanish, and Catalan, while the Arabic version is on its way. The user may also decide whether they would like to remain signed in to \uhelp\ or not. Under the ``Profile'' sub-view (Figure~\ref{fig:profile}), the user may edit its name, email, and password, as well as manage its list of relatives and locations. Saving one's most referenced relatives and locations simplifies asking for help. Under the ``Privacy'' sub-view, the user may decide who is allowed access to their personal phone number. As illustrated earlier in Section~\ref{sec:requestsview}, requesters and volunteers may contact each other by chat or phone. While chatting via \uhelp\ does not reveal one's phone number, making a phone call does, which raises privacy concerns. As such, there are three settings to choose from. In the first, the phone number will be hidden from everyone (even friends), and in this case no one can call them via \uhelp. In the second, one's phone number will be accessible either by requesters that one volunteers to their private requests or volunteer that respond to their private requests, and in this case, the requester and the volunteer can contact each other via phone calls. Note that by limiting requests to private ones (see Section~\ref{sec:helpview} on the difference between private and public requests) this ensures that the phone number is only shared with people in one's own social network, that is, people that are reachable through their friends (e.g. friends of friends, or friends of friends of friends, and so on). In the third privacy setting, one's phone number will be accessible either by any requester that one volunteers to help or any volunteer that respond to their requests (whether the requests where private or public), and in this case, the requester and the volunteer can contact each other via phone calls. 

Last, but not least, we note that the user can sign out of \uhelp\ in the Settings view. 

\subsection{Functionality}\label{sec:fsm}
\begin{figure}
\centering
\includegraphics[width=0.9\textwidth]{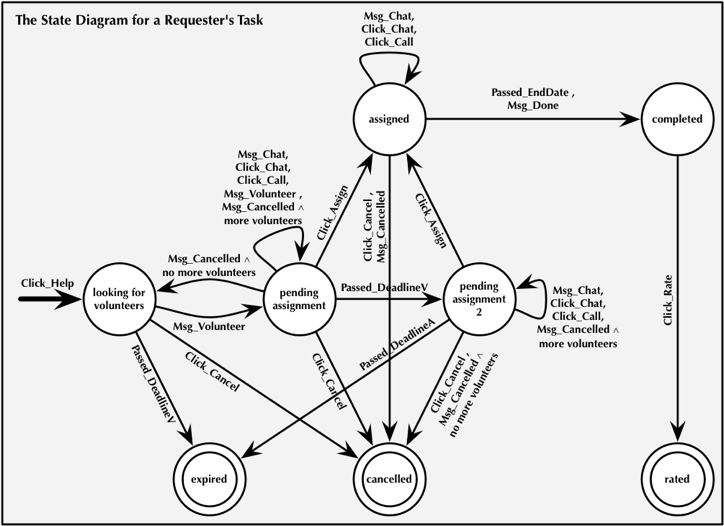}
%\caption{The requester's state diagram}
%\label{fig:fsm_requester}
%\end{figure}
%
%\begin{figure}
%\centering
\includegraphics[width=0.9\textwidth]{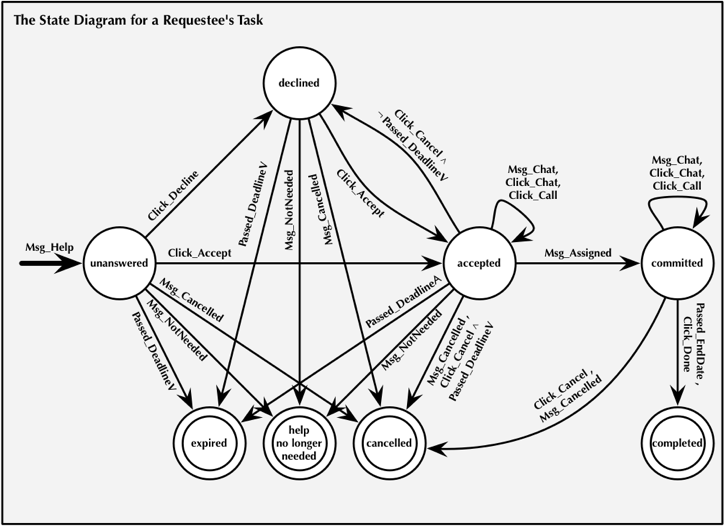}
\caption{The state diagrams of the requester and the requestee, respectively}
\label{fig:fsm}
\end{figure}

In this section, we define the functionality of the \uhelp\ app. This is illustrated by the state diagrams of Figure~\ref{fig:fsm}. Note that there are different state diagrams for the requester and the requestee (or potential volunteer). 

\subsubsection{The functionality with respect to the requester}
The state diagram of the requester shows eight different states. Anyone can ask for help at any point in time. When a requester clicks the help button (``Click\_Help"), the request is created and added to the requester's list of requests, its state is initially set to \emph{looking for volunteers}, and the flooding algorithm (Section~\ref{sec:flooding}) is called to propagate the request appropriately. As soon as one person volunteers (specified by receiving the message ``Msg\_Volunteer''), the state of the request changes to \emph{pending assignment}. At this point, more people may still volunteer (``Msg\_Volunteer'' at state \emph{pending assignment}), without changing the state. Every time someone volunteers, they are added to the set of volunteers for that specific request. At the \emph{pending assignment} state, the requester may also select a volunteer (``Click\_Assign''), %\footnote{This is in fact a two way handshake, but we keep it simple here. Though not sure how the two way handshake helps. For instance, what if the OK message was not received by the requster?! Then it is stuck at assigned. Maybe then it waits till deadline and move to closed?! Deadline must anyway be added to all non final states in case they get stuck there. To be discussed.}
moving the state to \emph{assigned}. If the task has been assigned to a volunteer, then the volunteer is expected to carry out that task in time. The state changes to \emph{completed} either when the deadline to perform the task passes (``Passed\_EndDate'') or when the volunteer marks the task as done (``Msg\_Done''), which is only allowed before the deadline. At this point, the requester may rate the volunteer's performance (``Click\_Rate''), which calls the trust model to update the trust values, and the state changes to the final state \emph{rated}.
% NAR XXX In the current implementation, we can only rate our friends.

We note that there are several deadlines used to ensure the state eventually moves to a final one. For instance, there is a deadline for volunteering, and as soon as this deadline passes (``Passed\_DeadlineV''), if the task has not been assigned to a volunteer yet, then the state changes either to \emph{pending assignment 2} when there are volunteers to choose from (that is, the deadline passes when at the state \emph{pending assignment}), or the final state \emph{expired} when there are no volunteers to choose from (that is, the deadline passes when at the state \emph{looking for volunteers}).  When at the state \emph{pending assignment 2}, the requester is once again asked to choose a volunteer, as no more people will be volunteering. The requester may select a volunteer at this state (``Click\_Assign''), again, moving the state to \emph{assigned}. However, we note that there is another deadline for selecting a volunteer. If the requester fails to select a volunteer in time (``Passed\_DeadlineA''), then the state will change to the final state \emph{expired}. 

As for communication between the requester and volunteers, we note that the requester may chat (``Click\_Chat'' for sending a chat message and ``Msg\_Chat'' for receiving a chat message) or even call a volunteer (``Click\_Call'', if the privacy settings permit calls) at any state when a volunteer exists and the task has not been completed yet (that is, states \emph{pending assignment}, \emph{pending assignment 2}, and \emph{assigned}). Performing such actions do not change the state of the request. 

At any point in time, the requester may also cancel the request, except after it has been completed, of course. As such, a cancel action (``Click\_Cancel'') is permitted at all non-final states, except for the \emph{completed} state. This action changes the state to \emph{cancelled}, which is a final state. Volunteers are also allowed to cancel, or withdraw from volunteering (``Msg\_Cancel''). Again, this is possible at most non-final states, except for the \emph{completed} state and the \emph{looking for volunteers} state (that is, a volunteer cannot cancel before volunteering). When a volunteer cancels, first, they are removed from the set of volunteers for this specific request. If they cancel after the task has been assigned to them (states \emph{assigned} and \emph{assigned 2}), then this results in changing the state of the request to the final state \emph{cancelled}. In other words, the request is cancelled because it is complicated at this stage to go back and look for volunteers once again after they have been informed that their help is no longer needed. But if they cancel before the task is assigned to a volunteer (states \emph{pending assignment} and \emph{pending assignment 2}), then the state does not change if there are other volunteers available. However, if no other volunteers are available, then the state changes either to \emph{looking for volunteers} when the deadline to volunteer has not passed yet, or to \emph{cancelled} when the deadline to volunteer has passed. 

Last, but not least, we note that the flooding algorithm is not only called when a help request is made (``Click\_Help''), but in two other cases as well. First, when a request is cancelled by the requester before it is assigned to a volunteer (``Click\_Cancel'' at states \emph{looking for volunteers}, \emph{pending assignment}, and \emph{pending assignment 2}). In this case the flooding algorithm is asked to propagate a cancellation message so that all requestees are aware of this action. Also, when a volunteer is selected, the flooding algorithm is also called to inform all requestees (whether they already responded or not) that their help is no longer needed (``Click\_Assign'' at states \emph{pending assignment} and \emph{pending assignment 2} is coupled with sending the message ``Msg\_NotNeeded'' to other requestees, as illustrated by Section~\ref{sec:requesteeFSM}). 

\subsubsection{The functionality with respect to the requestee}\label{sec:requesteeFSM}
The state diagram of the requestee also has eight states and it complements that of the requester. From the requestee's point of view, when the requestee is asked for help (``Msg\_Help''), the request is created and added to its list of requests, and its initial state is set to \emph{unanswered}. At this point, the requestee may either decline (``Click\_Decline''), moving the state to \emph{declined}, or accept (``Click\_Accept''), moving the state to \emph{accepted}. The requestee is then given the chance to change its mind. So when at the \emph{declined} state, it may still accept (``Click\_Accept''), moving the state to \emph{accepted}. And similarly, when at the \emph{accepted} state, it may still cancel (``Click\_Cancel''), though the new state will depend on the deadline to volunteer. If the deadline to volunteer has not passed yet (``$\neg$Passed\_DeadlineV''), then the state moves from \emph{accepted} to \emph{declined}, giving the requestee the chance of changing its mind and accepting again. However, if it cancels when the deadline to volunteer has already passed (``Passed\_DeadlineV''), then the state moves from \emph{accepted} to the final state \emph{cancelled}. Note that if the deadline to volunteer passes (``Passed\_DeadlineV'') when at any of the \emph{unanswered} or \emph{declined} states, then the state moves to the final state \emph{expired}. 

When at the \emph{accepted} state, the requestee awaits to be selected. If the requester chooses its volunteer, the requestee is informed. It will either receive a message informing it that it is the selected volunteer (``Msg\_assigned''), moving the state from \emph{accepted} to \emph{committed}, or receive a message informing it that its help is no longer needed (``Msg\_NotNeeded''), moving the state from \emph{unanswered}, \emph{declined}, or \emph{accepted} to the final state \emph{help not needed}.  Also recall that there is a deadline for requesters to select volunteers in time. If this deadline passes without receiving any message from the requester about the selection of volunteers (``Passed\_DeadlineA''), then the state moves from \emph{accepted} to the final state \emph{expired}. Finally, when at the \emph{committed} state, the requestee is expected to fulfil its commitment and perform the task it has volunteered for. The request moves to its final state \emph{completed} either when the requestee marks the request as completed (``Click\_Done''), or when the deadline to perform the task is reached (``Passed\_EndDate'').

As mentioned earlier, the requester may cancel the request at any time, before the request is marked as completed. In other words, a requester's cancellation action (``Msg\_Cancel'') may happen at any non-final state (\emph{unanswered}, \emph{declined}, \emph{accepted}, and \emph{committed}), resulting in moving that state to the final state \emph{cancelled}. The requestee may also cancel at a state where it is considered a volunteer (that is, states \emph{accepted} and \emph{committed}). If it cancels when it is the selected volunteer (state \emph{committed}), then this moves the state to the final state \emph{cancelled}. This is because going back and finding other volunteers is complicated at this stage. Cancelling before being selected (state \emph{accepted}), has already been discussed above. 

Last, but not least, recall that the requestee may chat (``Click\_Chat'' for sending a chat message and ``Msg\_Chat'' for receiving a chat message) or even call a requester (``Click\_Call'', if the privacy settings permit calls) at any state where it is considered a volunteer (that is, states \emph{accepted} and \emph{committed}). Performing such actions do not change the state of the request. 

\subsection{Architecture}\label{sec:architecture}
This section provides a brief introduction to the \uhelp\ architecture. The \uhelp\ architecture is divided into two main components: (1) the application that is downloaded and used by users; and (2) the server that supports user interactions.

The \uhelp\ application was developed via the Ionic framework,\footnote{\url{http://ionicframework.com}} which allows developing an application as a web application that would additionally be automatically translated into iOS and Android applications. This allowed us to build a cross-platform application in one go: we wrote the code once, and its compilation resulted in different application versions for different operating systems. We also made use of the Apache Cordova platform,\footnote{\url{http://cordova.apache.org}} for developing plugins that map a device's components to plugins in javascript, which were then used on the Ionic framework. This was basically needed to access necessary device features, such as contacts, notifications, etc., using a javascript call.

On the other hand, the server was implemented as a Java program that integrated different technologies, to support the necessary user interactions. %to provide the web services that the application needs. 
Specifically, we have used: (1) \emph{Eclipse Jetty}, as the server that manages http requests; (2) \emph{JAX-RS with Jersey}, to define and link http requests to their respective web services; (3) \emph{JAXB with Eclipse link}, to help convert the data model from/to XML or JSON; (4) \emph{JPA with Eclipse link}, to map the Java object to tables in a data base and vice-versa; %On the running version we link to a MySQl data base; 
(5) \emph{AKKA}, to provide the peer infrastructure needed for the flooding algorithm; (6) \emph{Apache shiro}, to authenticate and manage users' sessions; (7) \emph{notnoop/java-apns library}, to send push notification to the iOS devices; and (8) \emph{YouCruit/gcm-server library}, to send push notification to Android devices.

\subsubsection{Current implementation and future work}\label{sec:future}
The current implementation of \uhelp\ (version 2) is a simplified implementation of the model presented in Section~\ref{sec:technologies}; to be more precise, the semantic similarity and trust models presented earlier have been simplified. In our current implementation, the semantic similarity model is currently a binary model, where requests are considered either similar (with semantic similarity $1$) or not (with semantic similarity $0$). The trust model is a simplified model that only considers taxonomies. Ongoing work is currently focusing on implementing our more elaborate semantic similarity and trust models, with the extension of sharing ratings. 

Additionally, future work shall take into consideration the evolution of meronomies and taxonomies. We believe meronomies and taxonomies are community dependent: the tasks that interest one community might not interest another. As such, we suggest future work to allow for more dynamic meronomies and taxonomies, as well as allow community members to update these meronomies and taxonomies on the fly. 

Finally, we note that the current implementation is a centralised implementation. We plan to decentralise \uhelp\ in our future work. Our ultimate aim is to achieve a fully distributed platform. To achieve this, we plan to make use of Ethereum~\cite{buterin2014ethereum,wood2014ethereum}, an open-source distributed computing platform that is based on blockchain. We also note that when the platform is decentralised, then it will be important to take breaks in connectivity into account. To address this issue, we plan to make use of time limits, after which it is assumed that any node that has not answered was unreachable. This allows the algorithm to deal with users whose cellphones are switched off, or are unreachable for other reasons. Additionally, at a lower level, the messaging protocol could detect such failures and it can then resend the message when the receiving node is back online.

\section{Use case: a community of single parents}\label{sec:usecase}
The \uhelp\ application was both designed and evaluated with the support of the Catalan federation of single parent families.\footnote{\url{http://www.familiesmonoparentals.cat}}\footnote{The use case and its progress was announced on the blog of the Catalan federation of single parent families, which is available in the Cataln language at: \url{https://somunafamilia.wordpress.com/2016/02/19/les-mares-proven-lapp-mobil-uhelp/} and \url{https://somunafamilia.wordpress.com/2016/03/07/les-mares-comencen-la-prova-pilot-de-lapp-uhelp/}} At the beginning of our implementation, we had a meeting with 18 selected volunteers from the single parents community in Barcelona to get their feedback on our \uhelp\ idea.  After explaining the basic idea of how volunteers are searched for using the trust and friendship levels, the feedback was positive and very welcoming to our proposed intelligent volunteer search. To them, this implied an efficient alternative to existing methods that required either more manual work (namely, creating groups in Whatsapp and handpicking the recipients for some request) or less privacy (broadcasting a request to an already existing group where some members of that group are not considered adequate to receive their request). The volunteering parents also proposed some additional features that our implementation took into consideration, such as usability and privacy features as well as adding new tasks (the general ``Help me'' request as well as the ``Get me something'' and ``Substitute me'' requests were suggested by our user volunteers). And it was the user volunteers who required associating friendship and trust levels to tasks. 

After implementing \uhelp\ and making it available on Google Play and Apple Store, we revisited our volunteers. They were asked to download the app in our presence and play with it to see whether they had any difficulties. Despite the initial positive excitement and requirements, understanding the functionality was not straight forward this time. They essentially needed reminding of the functionality to understand how requests are propagating, as the application was not intuitive enough. 

The volunteers were then asked to use the application for a month. To help us analyse the \uhelp\ use case further, we had implemented a KPI dashboard which essentially visualised the evolution over time for a number of relevant KPIs (key performance indicators), such as the number of registered users, the number of signed in users, the average number of friends per user, the number of requested tasks, completed tasks, expired tasks, and so on. 

While the results of this initial experiment with the single parents community highlighted the need for a much more intuitive user interface, the community remains interested and excited about the \uhelp\ application. Further improvements continue to be developed and tested with this community, as we illustrate shortly in Section~\ref{sec:conclusion}.

\section{Conclusion}\label{sec:conclusion}
This article has presented \uhelp, a social networking app that provides a %fully distributed 
platform for building and maintaining a local community of people helping each other with their day-to-day tasks. It allows for one to find a trustworthy volunteer by automatically searching his social network. 

What distinguishes \uhelp\ from existing social network applications is its intelligent search for volunteers. Existing social network platforms allow one to broadcast their request either publicly or privately for their group of friends, or even create a group of manually handpicked friends for discussing a certain topic. \uhelp, on the other hand, crawls one's own social network looking for suitable volunteers. It essentially propagates a help request along one's own social network through a flooding algorithm that forwards a request from one node in one's social network to another based on the satisfaction of the required levels of trust and friendship.

While existing platforms may be adequate for some requests, like finding help searching for children summer activities, \uhelp\ addresses the more sensitive and urgent tasks. For example, a parent who is running late at work and realises the last minute that he needs someone to pickup his child from school. \uhelp\ allows this person to efficiently and automatically search for and ask all appropriate potential volunteers in his social network.  

Last, but not least, \uhelp\ was both designed and evaluated with a real user community, a community of single parents in Barcelona. The parents' initial feedback was used in the design of the application, as illustrated earlier. And their feedback from using the application was highly valuable for the usability engineering, which is driving the ongoing work. 

The \uhelp\ app that is currently available online remains an initial prototype, and one of important ongoing work for the users is improving its usability. %In fact, the current online version implements a very simplified version of the suggested semantic similarity and trust models presented earlier. The semantic similarity model is currently a binary model, where requests are considered either similar (with semantic similarity $1$) or not (with semantic similarity $0$). The trust model is a simplified model that only considers taxonomies. Ongoing work is currently focusing on implementing our more elaborate semantic similarity and trust models, with the extension of sharing ratings. 
Additionally, and as illustrated earlier (Section~\ref{sec:future}, the current application is a centralised application that implements a very simplified version of the suggested semantic similarity and trust models of Section~\ref{sec:technologies}. Ongoing work is currently implementing our more elaborate semantic similarity and trust models, with the extension of sharing ratings. And future work will focus on providing a fully decentralised and distributed platform. We also plan to have dynamic meronomies and taxonomies that can be updated by community members themselves.  
Additionally, future work, which has already commenced, also takes into consideration building different communities, where a user may be a member of several communities, and may decide to which community to post a request. For instance, one parent may be a member of the parent-teacher association of her children's school, a member of her workplace community, and a member of the local skating community. 

New versions of the app that will implement these new features will continue to be tested with the single parents community in Barcelona, as well as additional new communities (such as existing time bank communities and communities supporting refugees in Europe, where volunteer work is central for both).  %Nevertheless, the main contribution of \uhelp\ remains centred around its AI-based intelligent volunteer search.

\section*{Acknowledgements}
This work is supported by the uHelp project (Spanish Ministry of Economy and Competitiveness, under grant number EUIN2015-62530).



\section*{References}
\bibliographystyle{elsarticle-num} 
\bibliography{references}


%% %% \bibitem{label}
%% %% Text of bibliographic item
%%
%%\bibitem{}
%%
%%\end{thebibliography}
\end{document}